\documentclass[final,5p,times,twocolumn]{elsarticle}

\usepackage[numbers]{natbib}

\usepackage[T1]{fontenc} 

\usepackage[utf8]{inputenc}
\usepackage{url}
\usepackage{graphicx}
\usepackage{esvect}
\usepackage{graphics}
\usepackage{natbib}
\usepackage{mathrsfs}
\usepackage{blindtext}
\usepackage{footnotehyper}
\makesavenoteenv{figure}
\usepackage[dvipsnames]{xcolor}

\usepackage[colorlinks=true,breaklinks=true]{hyperref}
\usepackage[normalem]{ulem}
\usepackage[utf8]{inputenc}
\hypersetup{allcolors=[rgb]{0.0 0.0 0.6},linkcolor=[rgb]{0.75 0.05 0.05}}
\usepackage{amsmath,amssymb}

\usepackage{epsfig}  
\usepackage{slashed}       
\usepackage{tikz}
\usepackage{color}
\usepackage{url}
\usepackage{multirow}
\usepackage{bm}
\usepackage{footmisc}

\usepackage{comment}
\usepackage{aas_macros}

\usepackage{lineno}

\usepackage{multirow}

\usepackage[normalem]{ulem}
\usepackage{color}

\definecolor{jzp}{rgb}{0.8, 0.33, 0.0}

\usepackage{soul}
\usepackage{supertabular}
\usepackage{tabularx}

\allowdisplaybreaks

\bibpunct{[}{]}{,}{n}{}{,}

\journal{Physics of the Dark Universe}

\begin{document}

\begin{frontmatter}

\title{The potential of diffuse Galactic Ridge neutrino measurements to constrain dark matter}

\author[inst1,inst2]{Jaume Zuriaga-Puig\corref{cor1}}
\ead{jaume.zuriaga@csic.es}
\cortext[cor1]{Corresponding author.}
\author[inst1,inst2]{Pedro De La Torre Luque}
\ead{pedro.delatorre@uam.es}
\author[inst3]{and Viviana Gammaldi}
\ead{viviana.gammaldi@ceu.es}

\affiliation[inst1]{organization={Instituto de Fisica Teorica UAM-CSIC, Universidad Autonoma de Madrid}, 
            addressline={C Nicolas Cabrera, 13-15}, 
            city={Madrid},
            postcode={28049}, 
            country={Spain}}
\affiliation[inst2]{organization={Departamento de Fisica Teorica, M-15, Universidad Autonoma de Madrid},
            city={Madrid},
            postcode={E-28049}, 
            country={Spain}}

\affiliation[inst3]{organization={Department of Information Technology, Escuela Politécnica Superior, Universidad San Pablo-CEU}, 
            addressline={Campus Montepríncipe, Boadilla del Monte}, 
            city={Madrid},
            postcode={28668}, 
            country={Spain}}

\begin{abstract}

We use the latest ANTARES Galactic Ridge neutrino measurements to investigate their implications for indirect dark matter (DM) searches. We consider both annihilating and decaying DM scenarios, spanning a wide range of masses and final states, and systematically compare the resulting neutrino fluxes with the expected astrophysical Galactic diffuse emission. Furthermore, we compare the results for different DM density profiles allowed by the observations, from spike and cuspy to cored profiles. We do so for the WIMP model-independent scenario and explore two more specific models: branons and very heavy sterile neutrinos, where a cold DM candidate arises naturally from the theory. We show the potential neutrino measurements in the Galactic Ridge for DM and make predictions for future neutrino observatories.

\end{abstract}


\end{frontmatter}

\section{Introduction}
\label{sec:intro}

Despite dark matter (DM) composing $\sim 84\%$ of the total matter content of the Universe \cite{Aghanim:2018eyx}, its nature is still one of the biggest mysteries in Physics. Although the existence of DM has been confirmed form a wide range of observations —e.g., rotation curves \cite{2021PDU....3200826B, McMillan2017}, the cosmic microwave background \cite{Aghanim:2018eyx} or the big bang nucleosynthesis \cite{2009NJPh...11j5028J}, among others— many DM particle models that can account for the observed relic abundance, $\Omega_\mathrm{DM}h^2 \simeq 0.12$, remain compatible with current constraints. Among all the candidates, the most widespread in the community are the Weakly Interacting Massive Particles (WIMPs), which can be thermally produced via the so-called freeze-out mechanism (for a review, see \cite{2024arXiv240601705C}). This kind of production mechanism implies that the DM particles can annihilate or decay, producing highly energetic Standard Model (SM) particles in the process, which, after propagating, create an imprint that could be potentially observed by gamma-ray, neutrino, and cosmic ray (CR) observatories.

The detection of the final particles from DM annihilation or decay falls under DM indirect searches. While extensive literature exists on this topic~\cite{Di_Mauro_2021, Acharyya_2021, 2012PhRvD..86j3506C, 2023JCAP...11..063Z, 2018EPJC...78..831A, 2023JCAP...10..003A, 2023PhRvD.108j2004A, 2025arXiv251100918T, 2023arXiv230804833J, ALBERT2020135439, 2025JCAP...03..058A, 2020arXiv200505109S, 2017EPJC...77..627A, 2016EPJC...76..531A, 2022PhRvL.129z1103C, 2016PhRvD..93j3009B, 2024PhRvD.109d3034A, 2023JCAP...12..038A, 2024PhRvL.133f1001C, 2025arXiv250909609R, 2016JCAP...02..039M, PhysRevLett.129.111101, 2022PhRvD.105d1301M, 2025arXiv250907982B}, no clear signal has been detected so far. However, despite the null result, competitive constraints have been set in the DM mass vs. annihilation cross-section/decay lifetime parameter space. Among the diverse astrophysical targets for these searches, the Galactic Center (GC) remains one of the most promising due to its proximity and high DM density, making it the expected source of the brightest signal. Yet, as an active region, distinguishing potential DM signals from various sources and backgrounds presents challenges~\cite{Di_Mauro_2021, Acharyya_2021, 2012PhRvD..86j3506C, PhysRevLett.129.111101}. A second potential challenge in the GC is the uncertainties on the inner shape of the DM density profile~\cite{2021PDU....3200826B, 2023JCAP...11..063Z, 2015A&A...578A..14C}, which can induce differences in the expected DM flux up to several orders of magnitude.

Indirect DM searches with neutrinos offer several advantages, including reduced astrophysical absorption and sensitivity to heavy DM candidates that are difficult to probe with gamma rays or charged CRs. The Galactic Ridge, a region spanning $|l| < 30^\circ$, $|b| < 2^\circ$ and characterized by a large DM column density and relatively well-controlled backgrounds, is therefore a compelling laboratory for such studies. Previous works have explored neutrino signals from DM in the GC and halo \cite{2014PhRvD..90d3004C}, demonstrating the complementarity between neutrino and gamma-ray observations. However, the potential of Galactic Ridge neutrino measurements has so far remained largely unexplored in this context.

High-energy neutrino astronomy is rapidly evolving into a mature observational field, opening a new window on the non-thermal Universe at TeV energies and beyond, while probing the relevant parameter space for DM: mass-lifetime (for decaying DM) or mass-annihilation cross-section (annihilating DM). Over the past decade, large-volume neutrino telescopes, such as IceCube \cite{2014PhRvD..90d3005A}, have established the existence of a diffuse astrophysical neutrino flux of extragalactic origin, while steadily improving their sensitivity to extended sources and diffuse emissions within our own Galaxy. Understanding the neutrino emission from the Milky Way has become a key objective, both as a probe of CR transport and interactions and as a potential window onto physics beyond the SM. Recently, the IceCube neutrino observatory offered the first conclusive measurement of the Galactic neutrino emission~\cite{IceCube2023_GalacticNeutrinos}. In turn, searches with the ANTARES neutrino telescope~\cite{Ageron2011_ANTARES, 2025PhR..1121....1A}, located in the Mediterranean Sea, found no significant detection of neutrinos from the Galactic plane~\cite{ANTARES:2025wvi}, due to the lower statistics collected. However, ANTARES has played a pioneering role in the exploration of the Galactic Ridge. Owing to its geographical location, the GC is observed by ANTARES through the Earth, strongly suppressing the background from atmospheric muons and providing a significant advantage over detectors located at the South Pole, such as IceCube, for this region of the sky. Exploiting this favorable geometry, ANTARES has reported the first hint of diffuse neutrino emission from the inner Galaxy, based on an extended region consistent with the Galactic Ridge \cite{2023PhLB..84137951A}. Although the statistical significance of the excess remains modest, these measurements represent a milestone in Galactic neutrino astronomy and provide a unique dataset to test both astrophysical and exotic scenarios.

Motivated by these developments, in this work, we use the latest ANTARES Galactic Ridge neutrino measurements to investigate their implications for indirect DM searches. We consider both annihilating and decaying DM scenarios, spanning a wide range of masses and final states, and systematically compare the resulting neutrino fluxes with the expected astrophysical Galactic diffuse emission (GDE). In doing so, we build upon earlier theoretical studies of Galactic neutrino emission and DM-induced neutrinos, while exploiting the unique observational capabilities of ANTARES. We further discuss the relevance of our results in light of recent and ongoing analyses by IceCube, which have been triggered in part by the ANTARES findings and already include dedicated template-based and segmented-region studies of the inner versus outer Galaxy. Besides, we discuss the possible systematics that can arise from the astrophysical emission and the DM density profile. We explore this approach for several DM candidates: in addition to the WIMP vanilla scenario (in a model-independent way, with an arbitrary annihilation/decay channel), we consider the particular case of branon DM \cite{2003PhRvL..90x1301C} and very heavy sterile neutrinos \cite{2014JHEP...07..044H}.

This paper is organized as follows. In Section~\ref{sec:neutrino_flux} we introduce the expected neutrino signal in the Galactic Ridge, both astrophysical and DM-induced. In Section~\ref{sec:Constraints}, we present our constraints to the WIMP vanilla scenario, both from annihilation and decay, and prospects for future neutrino telescopes considering a wide range of channels. In Section~\ref{sec:OtherDM} we explore other DM particle models, such as branons or very heavy sterile neutrinos. Finally, in Section~\ref{sec:conclusions}, we present our conclusions.

\section{Neutrino flux in the Galactic Ridge}
\label{sec:neutrino_flux}

The absence of an active galactic nucleus and of strong star-forming regions in the Milky Way results in a high-energy neutrino luminosity that is significantly lower than that expected for active galaxies contributing to the extragalactic neutrino background~\cite{Fang2024_NeutrinoDesert}. Galactic neutrino emission has only been tentatively detected in recent years, with fluxes well below those typical of actively star-forming galaxies. Indeed, observations of neutrinos from the Galactic plane support that the dominant contribution to the Galactic high-energy neutrino flux arises from the diffuse CR sea, resulting in comparatively reduced astrophysical foregrounds and thereby providing a more favorable scenario for DM searches. Recent theoretical and observational studies have reinforced this view, highlighting the challenges associated with detecting Galactic neutrinos against the dominant extragalactic background and atmospheric foregrounds. The innermost regions of the Milky Way —where gas densities, CR densities, and gravitational potentials peak— remain among the most promising targets for Galactic neutrino searches. In particular, the Galactic Ridge, encompassing the central molecular zone and the inner Galactic plane, is expected to be a bright emitter of neutrinos produced in hadronic interactions between CRs and interstellar gas.

\begin{figure}[t!] 
    \centering 
    \includegraphics[width=0.46\textwidth]{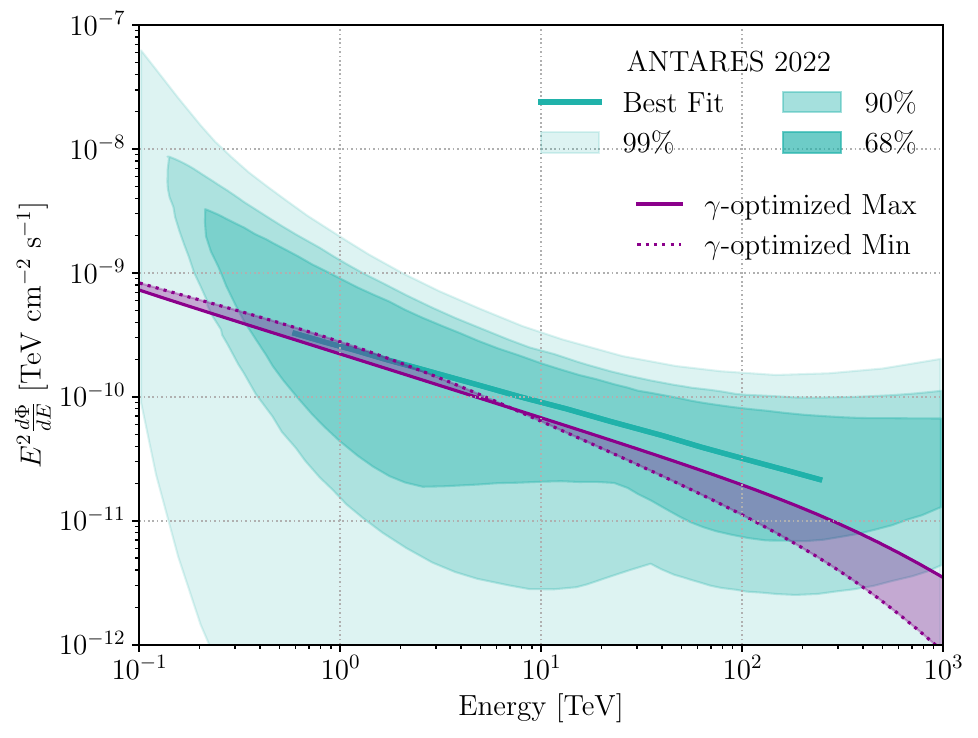}
    \caption{\footnotesize{ANTARES 2022 neutrino flux of the Galactic Ridge region (light blue bands) \cite{2023PhLB..84137951A}. As a comparison, we show in purple the GDE derived from the $\gamma$-optimized model (solid line for the Max model and dotted for Min model) \cite{DeLaTorreLuque2023_GammaPeV, DeLaTorreLuque2022_NeutrinoFlux}. The Galactic Ridge region is defined as $|l| < 30^\circ$, $|b| < 2^\circ$.}}
\label{fig:diffuse_only} 
\end{figure}

Even after 15 years (2007–2022) of data taken by ANTARES, the low statistics and high backgrounds make the measurements of the neutrino emission at the Galactic Ridge incredibly challenging. The Galactic neutrino emission is expected to be dominated by CR interactions occurring either right around the sources or later interacting with interstellar gas, roughly following, in any case, a smooth power-law behavior~\cite{AharonianDar2005, Gaggero2015}. Therefore, to perform this measurement and extract limits in the neutrino flux from this region, the ANTARES collaboration adopted power laws that are used as templates to reproduce the observed excess found in the Ridge in six independent energy bins~\cite{2023PhLB..84137951A, ANTARES:2025wvi}. To evaluate the systematic errors of assuming a given spectral index, they conservatively estimate the flux error in each bin when varying the spectral index $\alpha$ from 1 to 4, which is the reason for the resulting "peanut-like" shape of the error band (see Fig.~\ref{fig:diffuse_only}). This means that the confidence intervals that they provide are such that even sharp features (corresponding to spectral indices as low as $\alpha=1$) are accounted for in their measurement. For this reason, the upper limits in the neutrino flux at the Ridge may be used to put constraints on smooth signals originating from DM. In particular, we will explore the potential of these observations to probe different appealing DM scenarios, from the vanilla WIMP model to branons \cite{2003PhRvL..90x1301C, 2003PhRvD..68j3505C, 2006PhRvD..73c5008C} or very heavy sterile neutrinos \cite{2014JHEP...07..044H}. Interestingly, at the high masses that can be explored using these observations, electroweak corrections allow neutrinos to be produced even for the e$^+$e$^-$ final state  \cite{Cirelli_2011, Ciafaloni2010, 2021JHEP...06..121B, 2024JCAP...03..035A}, making the astrophysical neutrino measurements interesting for a variety of final states. Given the lack of detected sources in the region, we divide the total flux observed into two components: the GDE and the DM annihilation/decay flux.

\subsection{The neutrino Galactic diffuse emission}
\label{subsec:GDE}

As commented above, high-energy neutrinos in our Galaxy are predominantly produced through hadronic interactions of CRs with ambient matter. CRs (mainly protons and He) propagate through the interstellar medium, where they collide with hydrogen and helium nuclei, producing charged and neutral mesons. The decay of charged mesons generates neutrinos, while neutral mesons produce gamma rays. This process occurs both diffusively throughout the Galactic disk, as CRs scatter off gas clouds during their propagation, and locally near acceleration sites, such as supernova remnants, leading to localized neutrino emission that could be detected as point-like or extended sources in the sky. These two contributions —diffuse and source-like— together constitute the Galactic neutrino landscape.

Although current measurements of Galactic neutrinos are still statistically limited, they appear to be broadly consistent (within uncertainties) with standard models of CR interactions, as well as with the combined contribution from resolved sources and diffuse CR emission~\cite{Schwefer2022_CRINGE, Vecchiotti2023_GalacticNeutrinoFlux, DeLaTorreLuque:2025zsv, Marinos2025_GalacticNeutrinoGalprop, Ozlati_Moghadam_2026}. Interestingly, measurements by ANTARES indicate a relatively hard neutrino spectrum in the inner Galaxy compared to that inferred from local CR protons, in agreement with earlier hints from gamma-ray observations of the Galactic Ridge~\cite{2023PhLB..84137951A}. Such a spectral hardening may point to spatially non-homogeneous CR diffusion~\cite{Gaggero2015, TIBET, Cerri_2017}, potentially associated with enhanced turbulence, stronger magnetic fields, or CR re-acceleration processes in the central regions of the Milky Way. Alternatively, it could reflect a significant contribution from hadronic sources that dominate the diffuse neutrino flux and are strongly concentrated along the Galactic plane.
Similarly, the presence of a hard spectral component might hint at additional contributions beyond standard astrophysical processes. In this sense, the Galactic Ridge emerges as a particularly well-motivated target for indirect DM searches, whose annihilation or decay products could include high-energy neutrinos.

In recent years, increasingly sophisticated models of the GDE have been developed, incorporating spatially dependent CR diffusion, explicit source contributions, and constraints from multi-wavelength observations~\cite{Gaggero2015, Schwefer2022_CRINGE, Vecchiotti2023_GalacticNeutrinoFlux, DeLaTorreLuque:2025zsv}. Notably, several of these models are calibrated to reproduce gamma-ray data and are constructed independently of neutrino measurements.
In this work, we adopt as reference the models presented in Refs.~\cite{DeLaTorreLuque2023_GammaPeV, DeLaTorreLuque2022_NeutrinoFlux}, which have recently been confronted with the latest gamma-ray and neutrino data in Ref.~\cite{DeLaTorreLuque:2025zsv}. As shown in Fig.~\ref{fig:diffuse_only}, the neutrino flux predicted by these $\gamma$-ray–optimized GDE models exhibits remarkable agreement with the ANTARES Galactic Ridge measurements. This consistency supports a predominantly hadronic origin of the emission and, at the same time, lends credibility to the underlying assumption that the Ridge flux must dominate over other non-standard contributions.

\subsection{DM signals at the Galactic Ridge}
\label{subsec:DM_ridge}

The fact that the ANTARES Galactic Ridge is a region centered in the GC ($|l| < 30^\circ$, $|b| < 2^\circ$) makes it a compelling case for DM indirect detection studies. The GC has been extensively investigated with gamma rays, and in fact, it is in the GC where the best constraints  \cite{Di_Mauro_2021, Acharyya_2021, 2012PhRvD..86j3506C, PhysRevLett.129.111101} and detection prospects \cite{Di_Mauro_2021} have been obtained. For the case of neutrinos, the ANTARES Galactic Ridge observation only reports a $\gtrsim 2 \sigma$ hint detection when combining track and shower events, so, in order to be conservative, we will base our study comparing the upper limits with the expected astrophysical GDE and DM signal in the GC.

More specifically, in this work, we explore the possibility that the Galactic DM component is made by one out of three different candidates: a single candidate WIMP that annihilates or decays (by an effective s-wave process) into an arbitrary channel; the particular case of branon WIMP DM \cite{2003PhRvL..90x1301C} and very heavy sterile neutrinos \cite{2014JHEP...07..044H}. In all of these two cases, the annihilation/decay signals leave an imprint in both gamma rays, CRs and neutrinos. Assuming DM Majorana particles, the DM neutrino annihilation and decay flux is given by, respectively \cite{Cirelli_2011}:
\begin{equation}
\begin{split}
\frac{d\Phi^{\mathrm{ann}}_{\nu}}{dE}(E, \Delta\Omega) = \frac{\langle\sigma v \rangle}{8\pi m_{\mathrm{DM}}^2} \sum_{i}^{\mathrm{channels}} \mathrm{BR}_{i} \frac{dN^i_{\nu}}{dE} J(\Delta\Omega) \\
\frac{d\Phi^{\mathrm{dec}}_{\nu}}{dE}(E, \Delta\Omega) = \frac{1}{4\pi m_{\mathrm{DM}}\tau} \sum_{i}^{\mathrm{channels}} \mathrm{BR}_{i} \frac{dN^i_{\nu}}{dE} D(\Delta\Omega),
\end{split}
\label{eq:DM_flux_general}
\end{equation}
\noindent where $\langle\sigma v \rangle$ is the thermally averaged annihilation cross-section and $\tau$ is the lifetime of the DM particle. The summation of both formulas goes for the different SM particles created after the annihilation/decay of the DM particles. These SM particles are usually pairs of particles (fermions or bosons), like the $b\bar{b}$ or $\mu^+\mu^-$ channels, and are given by a set of branching ratios ($\mathrm{BR}_i$), defined such that $\sum_{i}^{\mathrm{channels}} \mathrm{BR}_i = 1$. During most of this work, we will always take the annihilation/decay to happen in only one channel, without the need to define branching ratios $\mathrm{BR_i}$, although for specific DM candidates, like branons or heavy sterile neutrinos, these branching ratios are directly computed from the theory. $\frac{dN^i_{\nu}}{dE}$ is the differential prompt neutrino emission produced at the source after the propagation, decay, or hadronization of the primary SM particles created. These final spectra can be computed with simulation codes, such as \texttt{PPPC} \cite{Cirelli_2011, Ciafaloni2010}, \texttt{HDMS} \cite{2021JHEP...06..121B}, or \texttt{CosmiXs} \cite{2024JCAP...03..035A}, and it depends mainly on the channel and DM mass $m_\mathrm{DM}$. In the literature, \texttt{PPPC} has been the main code for computing DM annihilation spectra in the usual cold DM WIMP masses ($ 5 \ \mathrm{GeV} \le m_\mathrm{DM} \le 140 \ \mathrm{TeV}$) \cite{Cirelli_2011, Ciafaloni2010, 2021JHEP...06..121B, 2024JCAP...03..035A, 2018RvMP...90d5002B}, although \texttt{HDMS} can be extended for higher DM masses ($m_\mathrm{DM} \ge 100 \ \mathrm{TeV}$). Recently, the \texttt{CosmiXs} code was presented with an updated simulation of the SM particle physics model, showing some differences in the prompt flux with respect to \texttt{PPPC}. We will use these three codes to compare our results in the case of the neutrino prompt flux. The \texttt{PPPC} tables have long served as a standard reference and have been progressively refined over the years. Given their widespread use in the community, we adopt \texttt{PPPC} as our baseline, while also comparing results with those obtained using alternative codes. Of particular importance is the accuracy of the \texttt{PPPC} tables near the resonant region (i.e., around the DM mass), which makes \texttt{PPPC} both better validated and more extensively explored than other currently available tables. This becomes especially relevant for the neutrino annihilation channels, where the spectra feature a clear peak.

Finally, the $J(\Delta\Omega)$ ($D(\Delta\Omega)$) is the J-factor (D-factor), also called astrophysical factors, and are defined as the integral of the DM density profile $\rho_\mathrm{DM}$ along the line of sight $l$ within a given solid angle $\Delta \Omega$:
\begin{equation}
\begin{split}
J(\Delta\Omega) = \int_{\Delta \Omega(\alpha_{\mathrm{int}})} \mathrm{d} \Omega \int_{l(\hat{\theta})_{\min }}^{l(\hat{\theta})_{\max }} \rho^{2}_{\mathrm{DM}}[r(l)] \mathrm{d} l(\hat{\theta}) \\
D(\Delta\Omega) = \int_{\Delta \Omega(\alpha_{\mathrm{int}})} \mathrm{d} \Omega \int_{l(\hat{\theta})_{\min }}^{l(\hat{\theta})_{\max }} \rho_{\mathrm{DM}}[r(l)] \mathrm{d} l(\hat{\theta})
\end{split}
\label{eq:Jfactor_general}
\end{equation}

Note that the J-factor (D-factor) and $\langle\sigma v \rangle$ ($\tau$) are, apart from being highly degenerate, one of the main parameters in the formula, since the normalization of the signal is directly proportional to them. Given that the inner Galactic DM density profile is highly uncertain, we use different DM density profiles in our analysis, from cuspy profiles with DM spikes \cite{2021PDU....3200826B, McMillan2017, Zhao:1995cp, PhysRevLett.83.1719, Sadeghian2013, PhysRevD.78.083506, doi:10.1142/S0217732305017391} to more cored profiles \cite{Burkert:1995yz}.

As a final remark, in order to reproduce the observed thermal relic abundance, $\Omega_\mathrm{DM}h^2 \simeq 0.12$ \cite{Aghanim:2018eyx}, the required annihilation cross-section must be of the order of $\langle \sigma v \rangle_\mathrm{th} \simeq 3\times 10^{-26}\ \mathrm{cm^{3}}\,\mathrm{s^{-1}}$, the so-called thermal relic cross-section\footnote{This holds in general if $5 \ \mathrm{GeV} \lesssim m_\mathrm{DM} \lesssim 100-140 \ \mathrm{TeV}$ and the annihilation cross-section is through an s-wave process in which there is only one DM particle candidate. For heavier masses, however, the usual freeze-out mechanism for DM can break unitarity, in which this value does not hold anymore \cite{2018RvMP...90d5002B}.}. On the other hand, for decaying DM, the current constraints are usually greater than the age of the universe \cite{2018EPJC...78..831A, 2023JCAP...10..003A, 2023PhRvD.108j2004A, 2025arXiv251100918T, 2023arXiv230804833J, 2022PhRvL.129z1103C, 2016PhRvD..93j3009B, 2024PhRvD.109d3034A, 2023JCAP...12..038A, 2016JCAP...02..039M}. Another main difference between the decay and annihilation flux is the maximum energy $E_\mathrm{max}$ allowed for the expected final neutrino flux. For the case of annihilation, this maximum energy corresponds to the DM mass $E_\mathrm{max} = m_\mathrm{DM}$, whereas for decaying DM is $E_\mathrm{max} = m_\mathrm{DM}/2$.

\begin{figure*}[t!]
    \centering 
    \includegraphics[width=0.9\textwidth]{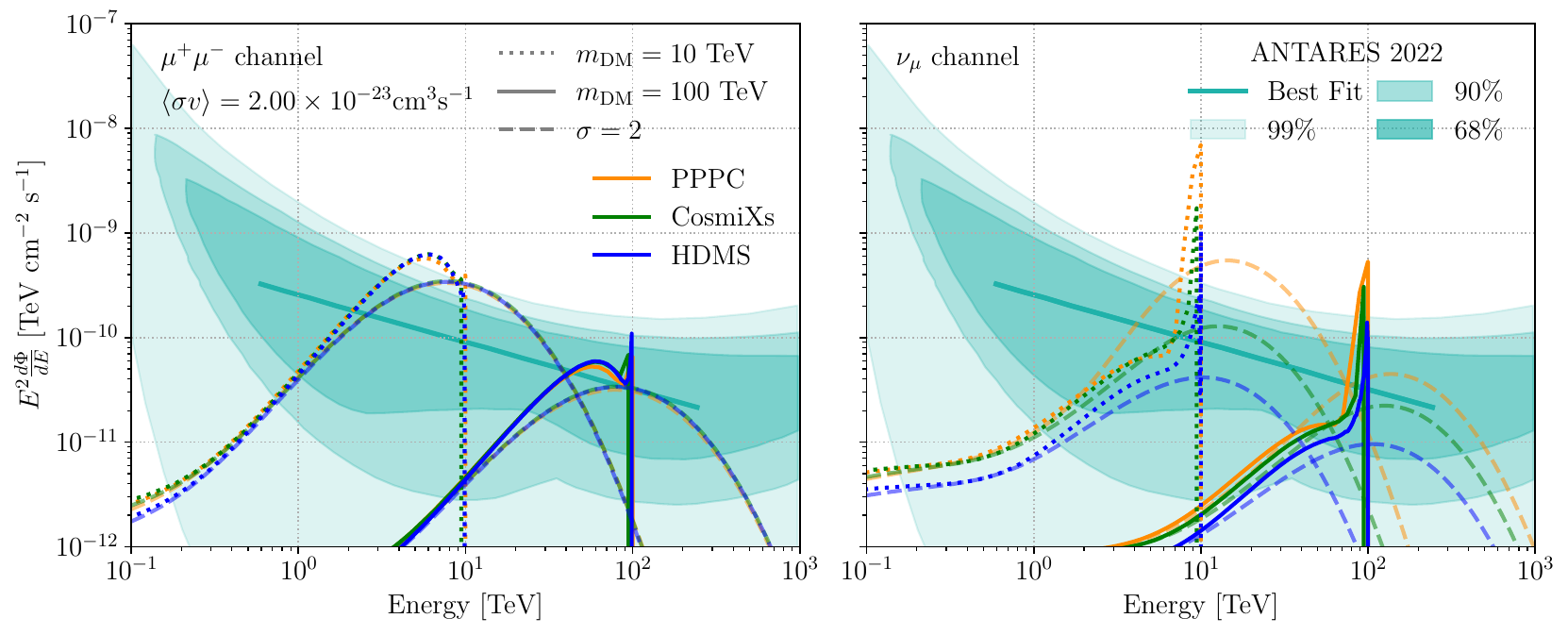}
    \caption{\footnotesize{ANTARES 2022 neutrino flux (light blue bands) compared with the expected neutrino flux in the Galactic Ridge from DM annihilation to the $\mu^+\mu^-$ (left panel) and $\nu_\mu$ channel (right panel). We show the spectra for two different masses ($m_\mathrm{DM} = 10$ TeV (dotted lines) and 100 TeV (solid lines)) and computed with three different codes (\texttt{PPPC} (yellow), \texttt{CosmiXs} (green) and \texttt{HDMS} (blue)). The DM spectra assume an NFW profile in both panels. In addition, we show with the dashed lines the fluxes after convolving them with the energy resolution kernel as Eq.~\ref{eq:energy_dispersion}.}}
\label{fig:DM_example_flux} 
\end{figure*}

Since many annihilation channels produce spectra with a pronounced peak in the expected signal, it is particularly important to convolve the flux with the ANTARES energy dispersion in order to avoid artificially mimicking a detection. With this aim, we model the energy resolution kernel on the observed flux $\left|\frac{d\Phi}{dE}(E_\mathrm{obs})\right|_\mathrm{obs}$ with the following equation:
\begin{equation}
    \left|\frac{d\Phi}{dE}(E_\mathrm{obs})\right|_\mathrm{obs} = \int_{-\infty}^\infty \left|\frac{d\Phi}{dE}(E)\right|_\mathrm{th} G(E; E_\mathrm{obs}, \sigma) dE,
    \label{eq:energy_dispersion}
\end{equation}
\noindent where $\left|\frac{d\Phi}{dE}(E)\right|_\mathrm{th}$ is the DM signal given by Eq.~\ref{eq:DM_flux_general} and $G(E; E_\mathrm{obs}, \sigma)$ is a lognormal gaussian defined by the parameters $\mu = E_\mathrm{obs} e^{(\log{\sigma})^2}$ and $\sigma$ as follows\footnote{Note that, with parametrization, the peak of the lognormal distribution corresponds to $E_\mathrm{obs}$.}:
\begin{equation}
    G(E; E_\mathrm{obs}, \sigma) = \frac{1}{E \log{\sigma} \sqrt{2\pi}} \exp{\left[-\frac{(\log{(E) - \log{(\mu)}})^2}{2 (\log{\sigma})^2}\right]}
    \label{eq:lognormal_function}
\end{equation}

Note that shower events, typically, report an energy resolution of $\sim 10-20\%$ \cite{2018ApJ...853L...7A, 2017EPJC...77..419A, 2024JCAP...08..038A}, while track events have an uncertainty of a factor $\sim 2$ \cite{2024JCAP...08..038A, 2006hep.ph...11032S}, which we approximate with a lognormal distribution. Thus, as the ANTARES Ridge flux combines both channels~\cite{2023PhLB..84137951A}, to compare the models with the data, we use $\sigma = 2$ pessimistically, corresponding to the worst-case resolution for track events. In \ref{ap:energy_resolution}, we illustrate the importance of the energy resolution for these studies by showing how the observed spectra vary when changing the $\sigma$ parameter.

In Fig.~\ref{fig:DM_example_flux} we show the comparison of the ANTARES Galactic Ridge observation (light blue region) with the expected DM signal for two representative annihilation channels (left panel for $\mu^+\mu^-$ channel and right panel for $\nu_\mu \bar{\nu_\mu}$) computed with two different DM masses (dotted line 10 TeV, solid line 100 TeV) and for three prompt emission simulations (\texttt{PPPC}\footnote{We use the \texttt{PPPC} prompt spectra with the inclusions of the electroweak corrections \cite{Cirelli_2011, Ciafaloni2010}.} (yellow), \texttt{CosmiXs} (green) and \texttt{HDMS} (blue)), assuming a thermal annihilation cross-section of $\langle\sigma v \rangle = 2.00 \times 10^{-23} \mathrm{cm^{3}} \mathrm{s^{-1}}$. Although in leptonic channels yield similar results, we can see that in the neutrino channels the main differences between the codes appear in the peak of the profile (i.e., at $E \sim m_\mathrm{DM}$), spreading up to more than 1 order of magnitude in the $m_\mathrm{DM} = 10$ TeV case. For this reason, we will compare our results obtained with the three codes. In addition, in the Figure, we show with the dashed lines the fluxes after convolving them as Eq.~\ref{eq:energy_dispersion}. The Figure clearly illustrates the importance of the energy convolution: after convolving the signal with the energy-resolution kernel, the spectral lines present in some annihilation channels are significantly smeared out, reducing the peak by almost 1 order of magnitude.

\subsubsection{Galactic DM density profile}
\label{subsec:DMProf}

We have stated in Eq.~\ref{eq:Jfactor_general} the importance of the DM density distribution profile in the DM signal. Because of the uncertainties in the internal shape of the Galactic DM profile, especially in the inner $r < 2$ kpc \cite{2021PDU....3200826B, 2023JCAP...11..063Z, 2015A&A...578A..14C}, there are several possible candidates. In general, the profiles commonly used for the GC are cuspy profiles such as the Navarro-Frenk-White (NFW) \cite{1996ApJ...462..563N, 1997ApJ...490..493N} and the Einasto \cite{1965TrAlm...5...87E} profiles:
\begin{equation}
    \rho_{\mathrm{NFW}} (r) = \frac{\rho_\mathrm{s}}{ (\frac{r}{r_{\mathrm{s}}}) (1 + \frac{r}{r_{\mathrm{s}}})^2} 
    \label{eq:NFW_profile}
\end{equation}
\begin{equation}
    \rho_{\mathrm{Einasto}}(r) = \rho_\mathrm{s} \exp{\left(\frac{-2}{\alpha} \left[\left(\frac{r}{r_\mathrm{s}}\right)^\alpha - 1\right]\right)},
    \label{eq:Einasto_profile}
\end{equation}
\noindent where $\rho_\mathrm{s}$ is the normalization parameter of the profile, $r_\mathrm{s}$ is the scale radius and $\alpha$ is the parameter that regulates the slope of the Einasto profile.

However, with the current observational data, cuspier profiles than an NFW profile are also allowed \cite{2021PDU....3200826B, McMillan2017, 2023JCAP...11..063Z, 2024A&A...692A.242G}. To also consider this possibility, we will also include two more profiles in our modelling. First, a generalization of the NFW profile (gNFW) that allows a variation in the internal slope $\gamma$, where the NFW case corresponds to $\gamma = 1$, although typically it can range between 0 and 1.5 \cite{2021PDU....3200826B, McMillan2017}:
\begin{equation}
    \rho_{\mathrm{gNFW}} (r) = \frac{\rho_\mathrm{s}}{ (\frac{r}{r_{\mathrm{s}}})^{\gamma} (1 + \frac{r}{r_{\mathrm{s}}})^{3-\gamma}} 
    \label{eq:gNFW_profile}
\end{equation}

Secondly, it has been proposed the presence of a DM spike in the inner $\sim 10 -20 \ \mathrm{pc}$ due to the adiabatic growth of the supermassive black hole SgrA* \cite{PhysRevLett.83.1719, Sadeghian2013}. This DM spike is the consequence of a big enhancement in the DM density profile, with the assumption that SgrA* has grown very slowly since its creation, always positioned in the center of the GC and without major mergers disrupting it during the last $t_\mathrm{BH}\sim 10 \ \mathrm{Gyr}$ \cite{2026arXiv260118896M}. This profile has the same gNFW form outside the spike radius $R_\mathrm{sp}$ (typically $\sim 10 -20 \ \mathrm{pc}$), but inside the spike the form is given by $\rho_{\mathrm{sp}}(r) \simeq \rho_{\mathrm{R}}(1-\frac{2 R_{\mathrm{s}}}{r})^3 (\frac{r}{R_{\mathrm{sp}}})^{-\gamma_{\mathrm{sp}}}$, where $\rho_{\mathrm{R}} = \rho_\mathrm{gNFW}(R_\mathrm{sp})$ and $\gamma_\mathrm{sp} = (9 - 2 \gamma)/(4 - \gamma) \in [2.25-2.4]$. However, under the assumption that DM is annihilating/decaying in the GC, there is a maximum value allowed for the DM density profile $\rho_\mathrm{sat} = m_\mathrm{DM} / (t_\mathrm{BH} \langle \sigma v \rangle)$ (for the case of annihilating DM). Once the profile reaches this value, then it follows a more relaxed behaviour $\rho_{\mathrm{DM}}(r) = \rho_\mathrm{sat} (\frac{r}{R_{\mathrm{sat}}})^{-1/2}$, with this final slope created when taking into account the velocity distribution of the  DM particles at those radii \cite{Vasiliev2007}. Using a relativistic derivation of the Adiabatic Spike, the analytical form is given by \cite{Sadeghian2013}:
\begin{equation}
  \rho_\mathrm{Spike} (r) = 
     \begin{cases}
       0 & r < 2 R_\mathrm{S}\\
       \frac{\rho_\mathrm{sp}(r) \rho_\mathrm{sat}(r)}{\rho_\mathrm{sp}(r) + \rho_\mathrm{sat}(r)}  &  2 R_\mathrm{S} \leq r \leq R_\mathrm{sp}\\ 
       \rho_\mathrm{gNFW} (r) & r \geq R_\mathrm{sp},\\
     \end{cases}
  \label{eq:Adiabatic_Spike_profile}
\end{equation}
\noindent where $R_\mathrm{S}$ is the Schwarzschild radius and the spike radius \mbox{$R_{\mathrm{sp}} = \alpha_{\gamma} r_{s} (M_{\mathrm{BH}}/ (\rho_s r_{s}^3))^{(1/(3- \gamma))}$}, with $M_\mathrm{BH} = 4.297\times10^6 \, \mathrm{M}_\odot$ the mass of SgrA* \cite{GRAVITY_2021} and $\alpha_\gamma$ is computed numerically \cite{PhysRevLett.83.1719}.

However, this idealistic formalism doesn't account for possible interactions between accreted DM particles and other elements of the Galaxy. To alleviate the assumptions, we use a modification of the Adiabatic Spike that includes the heating of the DM particles, flattening slightly the DM spike \cite{PhysRevD.78.083506, doi:10.1142/S0217732305017391}. This new profile, which we call Stars-Spike, has the same form as the Adiabatic Spike but with a flattening of the spike to a slope of $\gamma_\mathrm{sp} = 1.5$ inside the radius of influence of SgrA* ($r_\mathrm{b} = 2 \ \mathrm{pc}$). Even with these highly steep profiles, only observational constraints exist for the extreme Adiabatic Spike case, in which the overall slope gNFW profile that creates the spike is allowed up to ($\gamma \lesssim 0.6$) \cite{2023JCAP...11..063Z, 2024A&A...692A.242G, Lacroix_2018, GRAVITY_2021}. From a DM annihilation standpoint, the original Adiabatic Spike \cite{PhysRevLett.83.1719, Sadeghian2013} has also been ruled out by comparing the expected DM flux with the observed HESS data in the GC \cite{2023JCAP...11..063Z}, which further motivates using this approach.

On the other end of the spectrum, cored profiles are also allowed by observations, although they are not preferred \cite{2021PDU....3200826B, McMillan2017}. In any case, to also include in our analysis the possibility of a cored profile in the Galaxy, we will also consider the Burkert \cite{Burkert:1995yz} profile, which yields the smallest J-factor and D-factors:
\begin{equation}
\rho_{\mathrm{Bur}}(r) = \frac{\rho_\mathrm{s}}{(1 + \frac{r}{r_\mathrm{s}})(1 + (\frac{r}{r_\mathrm{s}})^2)} 
\label{eq:Burkert_profile}
\end{equation}

\begin{figure}[t!] 
    \centering 
    \includegraphics[width=0.46\textwidth]{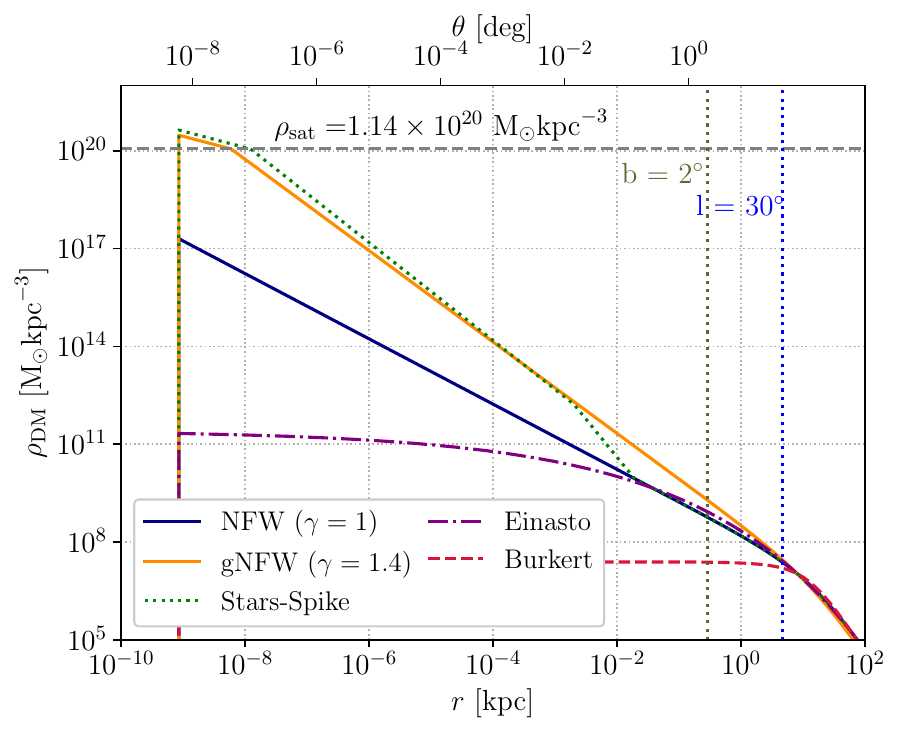}
    \includegraphics[width=0.46\textwidth]{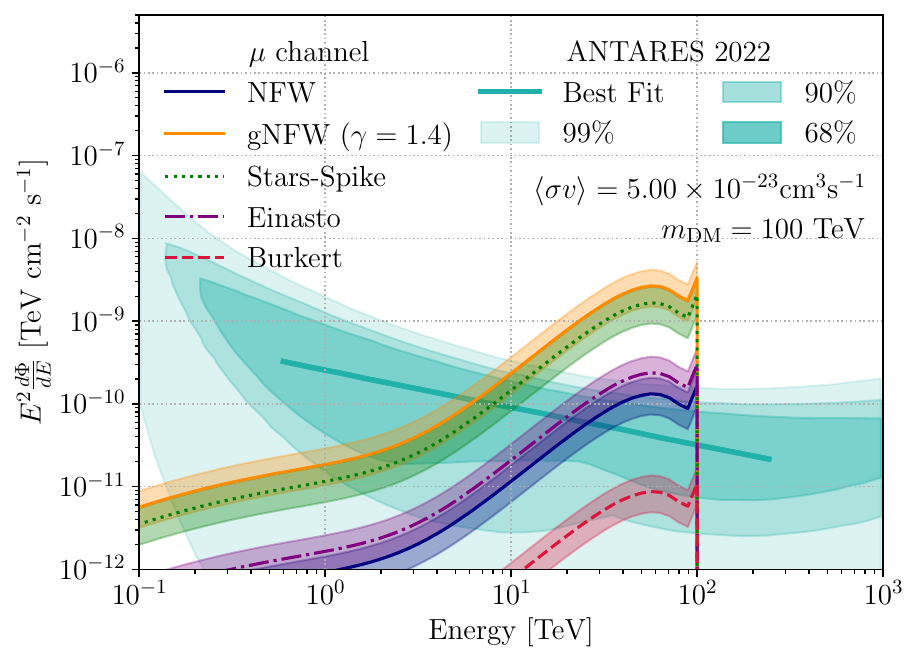}
    \caption{\footnotesize{Top panel: different Galactic DM density profiles used in this work, ranging from cuspy profiles (gNFW with $\gamma = 1.4$ (solid yellow line), Stars-Spike (dotted green), NFW (solid blue) and Einasto (dot-dashed purple)) and the Burkert cored profile (dashed red). The vertical dotted lines represent the boundaries of the ANTARES Galactic Ridge ($|b| \leqslant 2^\circ$, $|l| \leqslant 30^\circ$). Bottom panel: same color scheme as the top panel, we show a comparison between the neutrino DM fluxes (\texttt{PPPC}) and the ANTARES 2022 data for different profiles. Uncertainty bands correspond to rescaling of the J-factor for local densities between 0.3 and 0.5 $\mathrm{GeV cm^{-3}}$. The DM signal is computed from the a DM mass $m_\mathrm{DM} = 100$ TeV with $\langle \sigma v \rangle = 5.00 \times 10^{-23} \ \mathrm{cm^3 s^{-1}}$.}}
\label{fig:DM_profiles_range} 
\end{figure}

We show in the top panel of Fig.~\ref{fig:DM_profiles_range} the different range of DM density profiles (from cuspy profiles to cored) used in this work and that are allowed by the current Galactic rotation curves estimations \cite{2021PDU....3200826B} (solid blue line corresponds to the NFW profile, solid yellow for the generalized NFW profile ($\gamma = 1.4$), dotted green Stars-Spike, dot-dashed purple Einasto, and dashed red for the Burkert profile). We choose as a benchmark normalization of the profiles to have the same local DM density $\rho_\odot = 0.4 \ \mathrm{GeV} \, \mathrm{cm^{-3}}$ \cite{2021PDU....3200826B}, with the distance of the Sun to the GC taken as $R_\odot = 8.277$ kpc \cite{GRAVITY_2021}. We note that other values have been used in the literature from $8$ to $8.5$ kpc \cite{2021PDU....3200826B, McMillan2017, 2019A&A...625L..10G, 2013IAUS..289...29G, 2023MNRAS.519..948L}, but our results are not particularly sensitive to this value. For more information on the modeled profiles, in \ref{ap:app_density_profile}, we list the different parameters used in this work for all the DM density profiles and their corresponding J-factors and D-factors (Tabs.~\ref{tab:ROI_Dfactors} and \ref{tab:ROI_Jfactors}, respectively). In the bottom panel of Fig.~\ref{fig:DM_profiles_range}, we show how this range of DM density profiles translates into the expected annihilating flux of a DM particle of $m_\mathrm{DM} = 100 \ \mathrm{TeV}$ (computed with \texttt{PPPC}), compared with the ANTARES Galactic Ridge flux. The bands on the flux correspond to a change in the normalization $\rho_\odot$, which we leave to vary between $0.3$ and $0.5 \ \mathrm{GeV} \, \mathrm{cm^{-3}}$ to illustrate the dependence on this parameter.

\section{Probing vanilla WIMP with the Ridge observations}
\label{sec:Constraints}

Once the DM flux and the GDE are defined, we can combine both elements and compare them with the ANTARES Galactic Ridge flux to try to set constraints on the annihilation cross-section or lifetime parameters, i.e., the DM flux normalization. We show in this section the lower limits on the lifetime $\tau$ of decaying DM and the $\langle \sigma v \rangle$ upper limits. Regarding the DM density profiles, we choose the NFW as the baseline of this work, but note that if one wants to compute the constraints with other DM density profiles, they need to be computed with the corresponding D-factors or J-factors. In \ref{ap:app_density_profile}, we show the D-factors (Tab.~\ref{tab:ROI_Dfactors}), J-factors (Tab.~\ref{tab:ROI_Jfactors}) and the lifetime lower limits and $\langle \sigma v \rangle$ upper limits for the $\mu^+\mu^-$ channel with different DM density profiles: NFW, Einasto, Stars-Spike, gNFW ($\gamma = 1.4$) and Burkert. Given the wide dispersion of the D-factors and J-factors, we list in Tab.~\ref{tab:scaling_values_constraints} the corresponding rescaling values for the DM flux. We can see that a $\sim50\%$ uncertainty is present in the decay fluxes, whereas in the annihilating case it can get up to 1 or 2 orders of magnitude.

\begin{table}[h!]
    \begin{center}
    \resizebox{\columnwidth}{!}{
        \begin{tabular}{|c|c|c|c|c|}
        \hline
        \hline
\textbf{Case}                 &  Burkert  & Einasto  & Stars-Spike & gNFW ($\gamma = 1.4$) \\ 
\hline
\hline
Decay 
& 0.53 & 1.20 & 1.00 & 1.54 \\
\hline
Annihilation   
& 0.066 & 1.80 & 12.59 & 20.15 \\
\hline
\hline

        \end{tabular}
        }
        \caption{\footnotesize{Scaling values ($\equiv$ $\frac{d\phi}{dE}^i$/$\frac{d\phi}{dE}^{\mathrm{NFW}}$, where $i$ denotes a different DM profile) for the different DM density profiles presented in this work. For more information, see \ref{ap:app_density_profile} and Tabs.~\ref{tab:ROI_Dfactors} and \ref{tab:ROI_Jfactors}.}}
        \label{tab:scaling_values_constraints}
    \end{center}
\end{table}
\subsection{Constraints on decaying WIMPs: lifetime lower limits}
\label{subsec:lifetime_lims}

Although the ANTARES collaboration has mainly focused on exploring signals of WIMP annihilation in a variety of regions~\cite{2025PhR..1121....1A, Albert_2022, Albert:2019jpw, 2017PhLB..769..249A}, we show here that the Ridge measurements provide competitive limits for the case of decaying DM. To compute the lower limits, we base the analysis on the ANTARES Galactic Ridge upper limits ($90\%$ confidence level (C.L.)) as the maximum neutrino flux allowed in the region ($|l| < 30^\circ$, $|b| < 2^\circ$). Given that ANTARES only reports a $\sim 2\, \sigma $ detection, this approach allows us to be as conservative as possible without overshooting the observed data.

\begin{figure*}[t!] 
    \centering 
    \includegraphics[width=0.46\textwidth]{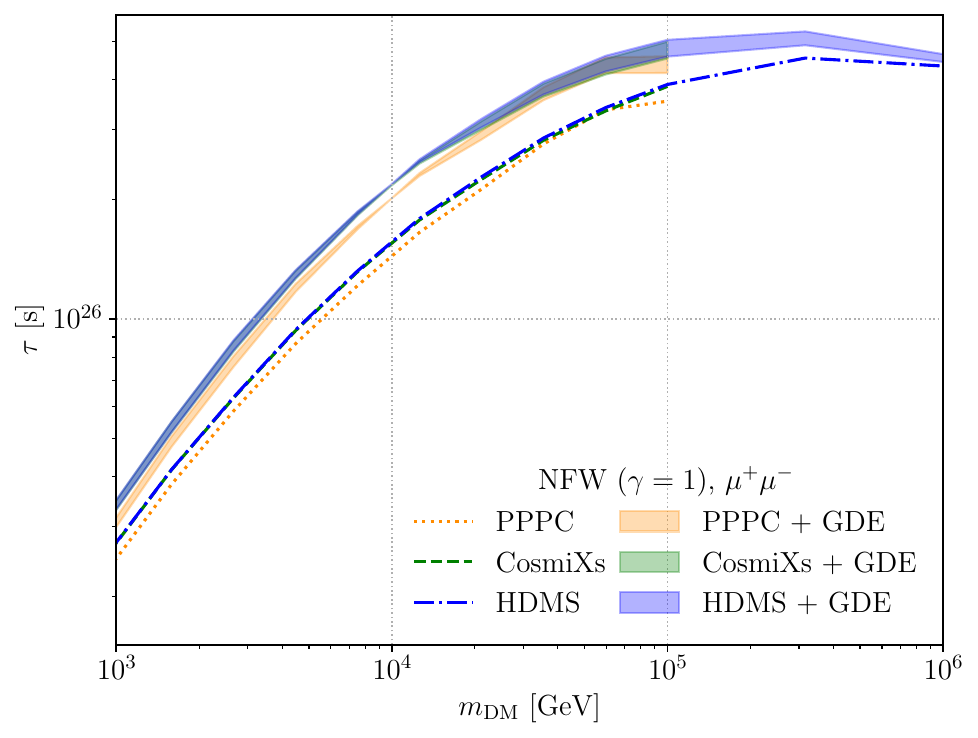}
    \includegraphics[width=0.46\textwidth]{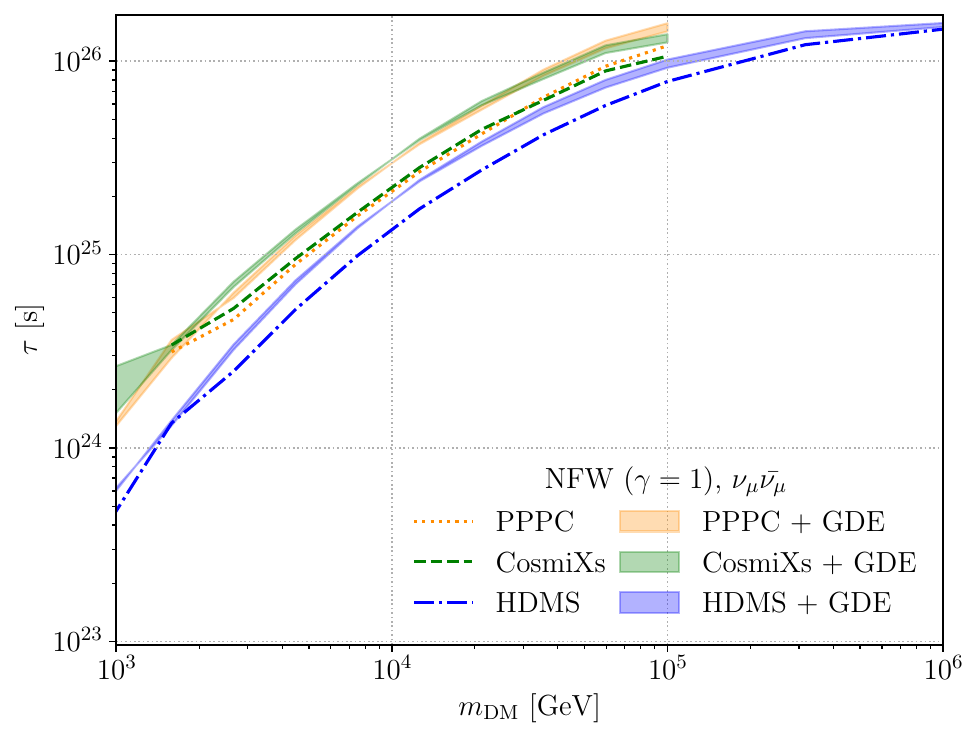}
    \includegraphics[width=0.46\textwidth]{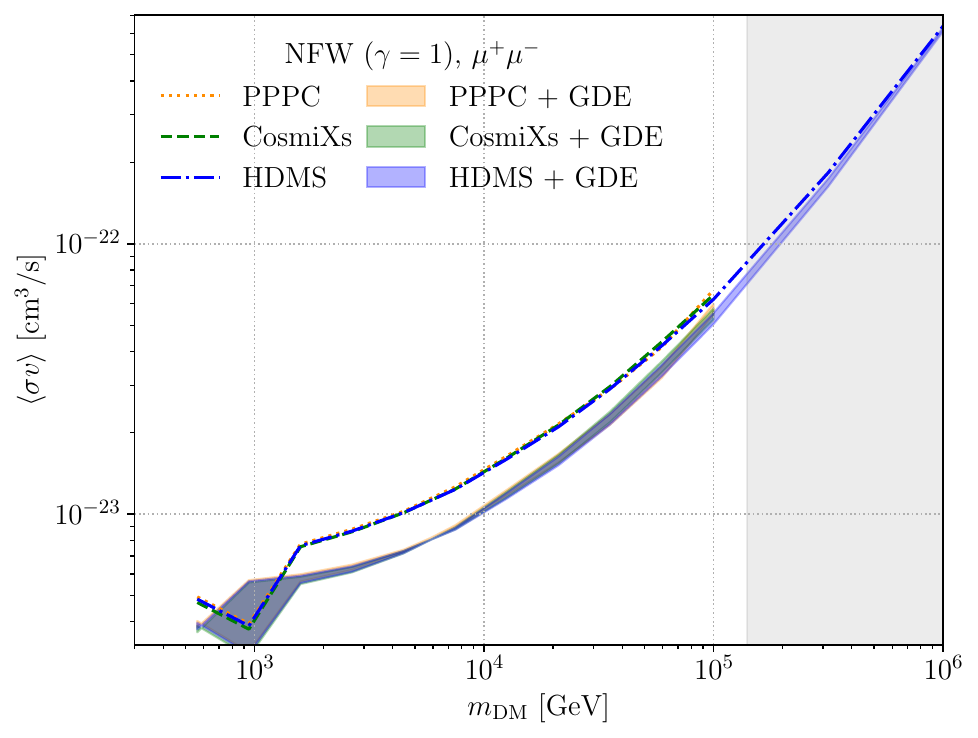}
    \includegraphics[width=0.46\textwidth]{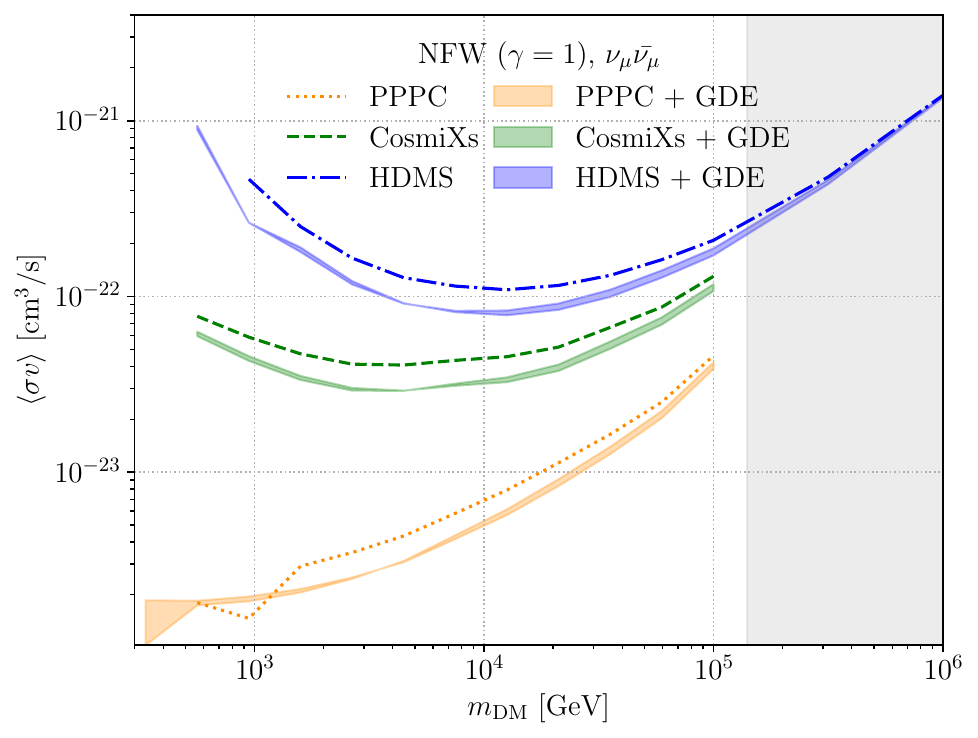}
    \caption{\footnotesize{$90\%$ C.L. lower limits on the annihilation decay lifetime $\tau$ (top row) and upper limits on the annihilation cross-section $\langle \sigma v \rangle$ (bottom row). The constraints are computed with the NFW profile for the $\mu^+\mu^-$ (left row) and the $\nu_{\mu}$ (right row) channels. As a comparison, we show the constraints computed from the \texttt{PPPC} (dotted yellow line), \texttt{CosmiXs} (dashed green) and \texttt{HDMS} (dot-dashed blue) codes. The colored bands correspond to the same constraints with the inclusion of the GDE $\gamma$-Opt model. The grey bands in the bottom row indicate when the unitarity limit breaks for thermal cold DM ($m_\mathrm{DM} \gtrsim 140$ TeV).}}
\label{fig:constraints_codes} 
\end{figure*}

Given the differences in the expected neutrino fluxes coming from the three different simulations presented in Fig.~\ref{fig:DM_example_flux}, we present the results always differentiating between \texttt{PPPC}, \texttt{CosmiXs} and \texttt{HDMS}. Also, the benchmark results are based on the standard NFW profile (Eq.~\ref{eq:NFW_profile}), but with the values presented in Tab.~\ref{tab:scaling_values_constraints}, all the constraints can be rescaled to other profiles. In the top row of Fig.~\ref{fig:constraints_codes}, we show the lifetime $\tau$ lower limits ($\mu^+\mu^-$ channel in the left panel and $\nu_\mu \bar{\nu_\mu}$ right panel) up to the $90\%$ C.L. for the three prompt spectra codes used in this work (yellow for \texttt{PPPC}, green \texttt{CosmiXs} and \texttt{HDMS} in blue). To distinguish the dependence when including the GDE model, we show the DM-Only case with the lines and the constraints including the GDE with the colored bands. As we can see, in the decay case, there is not much scatter between the three prompt spectra codes. This is because the main differences appear in the peak of the flux (when $E \sim m_\mathrm{DM}$), which corresponds to energies only possible in the annihilation case. Finally, we have noted that after convolving the DM signal with the ANTARES energy resolution (Eq.~\ref{eq:energy_dispersion}), the limits are less constraining by a factor of a few at most in some channels than in the perfect energy resolution case.

\begin{figure*}[t!] 
    \centering 
    \includegraphics[width=0.46\textwidth]{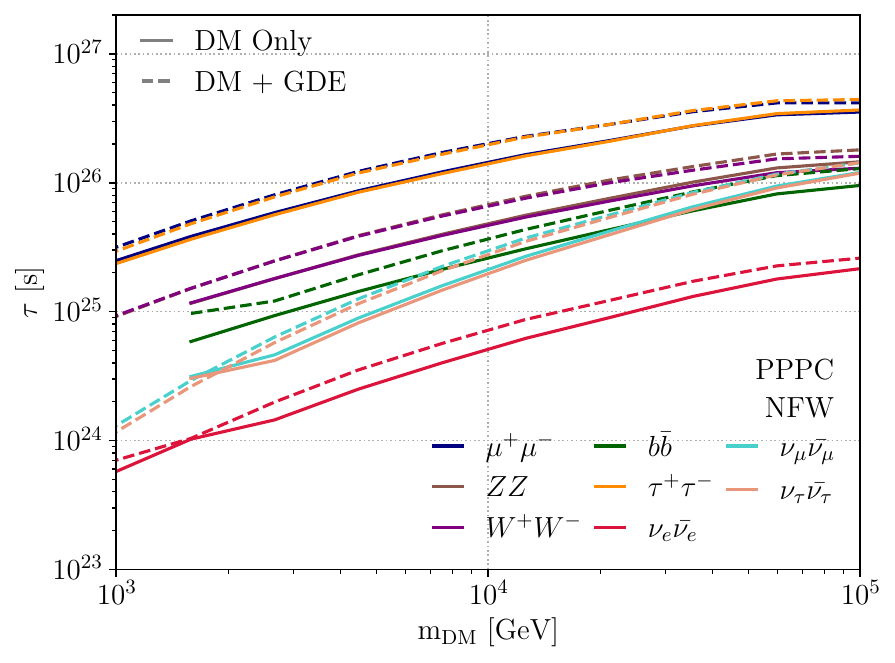}
    \includegraphics[width=0.46\textwidth]{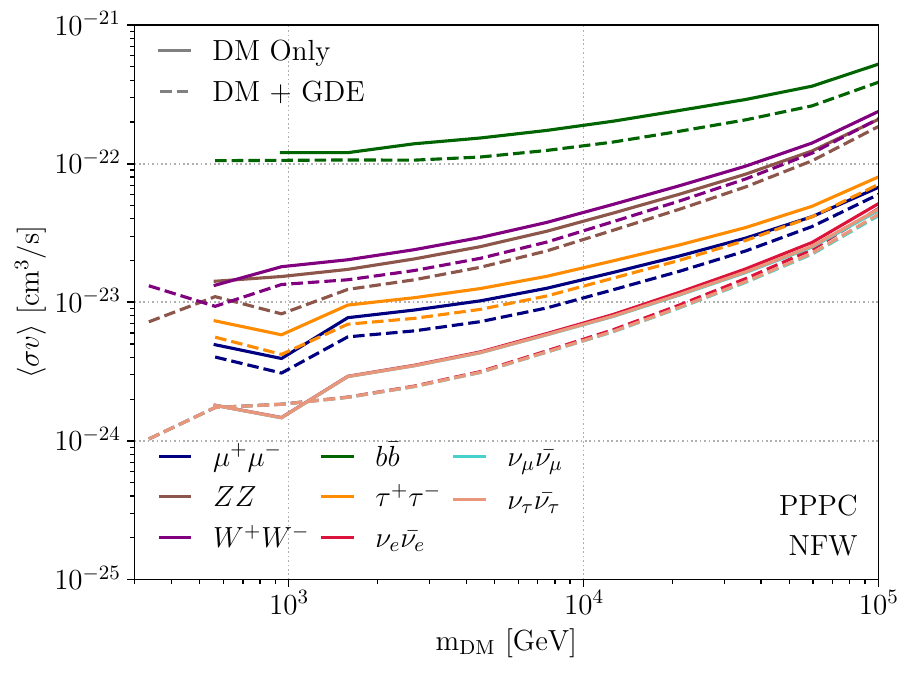}
    \caption{\footnotesize{$90\%$ C.L. lower limits on the annihilation decay lifetime $\tau$ (left panel) and upper limits on the annihilation cross-section $\langle \sigma v \rangle$ (right panel). We show the constraints with (dashed lines) and without (solid lines) the inclusion of the GDE model in the analysis. The constraints correspond to the \texttt{PPPC} simulation assuming the NFW profile. Note that, in the decay case, the $W^+W^-$ and $ZZ$ have very similar constraints.}}
\label{fig:all_constraints} 
\end{figure*}

Repeating this exercise with all the decay channels, we show in the left panel of Fig.~\ref{fig:all_constraints} the $90\%$ C.L. lower limits on the lifetime $\tau$ parameter, computed with the inclusion of the GDE emission (dashed lines) and DM-Only case (solid lines), all based on the \texttt{PPPC} simulation. The best results are given by the leptonic channels (such as $\mu^+\mu^-$ or $\tau^+\tau^-$). For a comparison with \texttt{HDMS} and \texttt{CosmiXs}, see \ref{ap:app_codes_prompt_emission}.

As a final comparison, in the top row of Fig.~\ref{fig:constraints_comparison}, we show our GDE+DM constraints with other constraints computed from gamma-ray \cite{
2022PhRvL.129z1103C, 2016PhRvD..93j3009B, 2024PhRvD.109d3034A, 2023JCAP...12..038A} (dashed lines) and neutrino \cite{2018EPJC...78..831A, 2023JCAP...10..003A, 2023PhRvD.108j2004A, 2025arXiv251100918T, 2023arXiv230804833J} (solid lines) data and from different targets\footnote{\label{fnote:CL_constraints}We note that not all the constraints are referred to the same C.L., as some refer to a $95\%$ and others to a $90\%$ exclusion region. Besides, not all the Galactic constraints share the same DM density profile.}. We do so for the $\mu^+\mu^-$ channel (left panel) and $\nu\bar{\nu}$\footnote{\label{fnote:nu_channel}We show the results for the $\nu\bar{\nu}$ channel, defined as an average of the three neutrino channels, although some of the results from the literature shown correspond to individual neutrino channels. But, for an easy comparison, we collect them in the same figure.} (right panel) channels, with the \texttt{PPPC} (yellow thick line) constraints. For the $\mu$ channel, our results are competitive with the literature, proving the potential of the Ridge measurements, although for the $\nu$ channel, our limits are more than an order of magnitude worse than the best IceCube's limits on the GC~\cite{2018EPJC...78..831A, 2025arXiv251100918T}. Similarly, in \ref{ap:app_tau_constraints_comparison} we show the same comparison with the $\tau^+\tau^-$ channel. This comparison is not exhaustive, as there are complementary probes of decaying DM, such as the isotropic gamma-ray background~\cite{2019JCAP...03..019B} or solar inverse-Compton searches~\cite{2026arXiv260400091D}, among others.

\subsection{Constraints on annihilating WIMPS: annihilation cross-section upper limits}
\label{subsec:app_annih_constraints}

We show in the bottom row of Fig.~\ref{fig:constraints_codes} the cross-section $\langle \sigma v \rangle$ upper limits for the different codes (yellow for \texttt{PPPC}, green \texttt{CosmiXs} and blue \texttt{HDMS}) and channels ($\mu^+\mu^-$ channel left panel, $\nu_\mu \bar{\nu_\mu}$ right panel), with (colored bands) and without (solid lines) the inclusion of the GDE. Given the differences between the peak of the neutrino fluxes between the three codes, the constraints can differ by even orders of magnitude in the neutrino channels. We also show the upper limits for the rest of the channels in the right panel of Fig.~\ref{fig:all_constraints} (with \texttt{PPPC}), reaching the best constraints for the neutrino channels but with barely any difference between the inclusion of the GDE (dashed lines) and without (solid lines). For a comparison with \texttt{HDMS} and \texttt{CosmiXs}, see \ref{ap:app_codes_prompt_emission}.

\begin{figure*}[t!] 
    \centering 
    \includegraphics[width=0.46\textwidth]{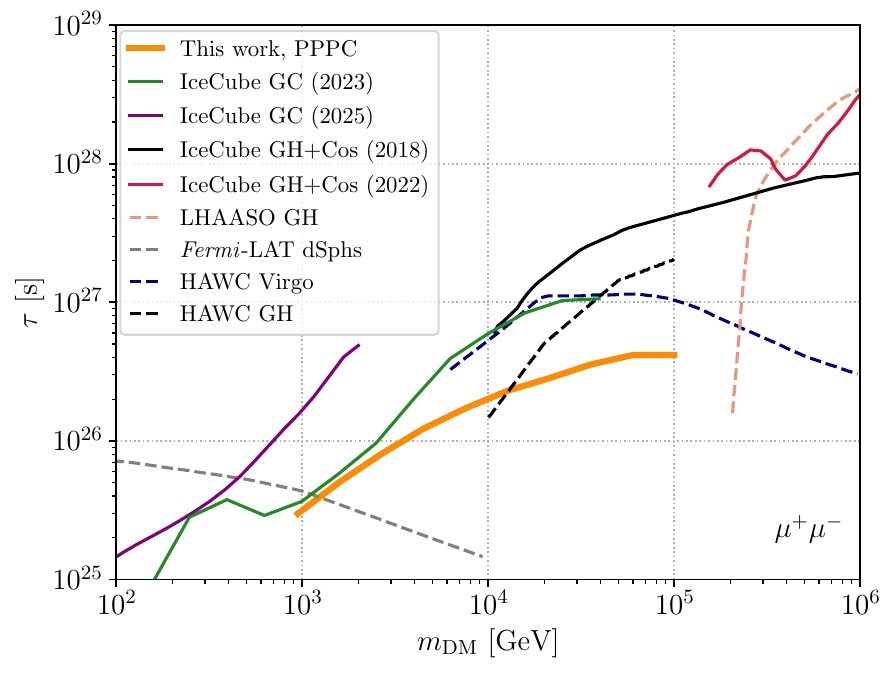}
    \includegraphics[width=0.46\textwidth]{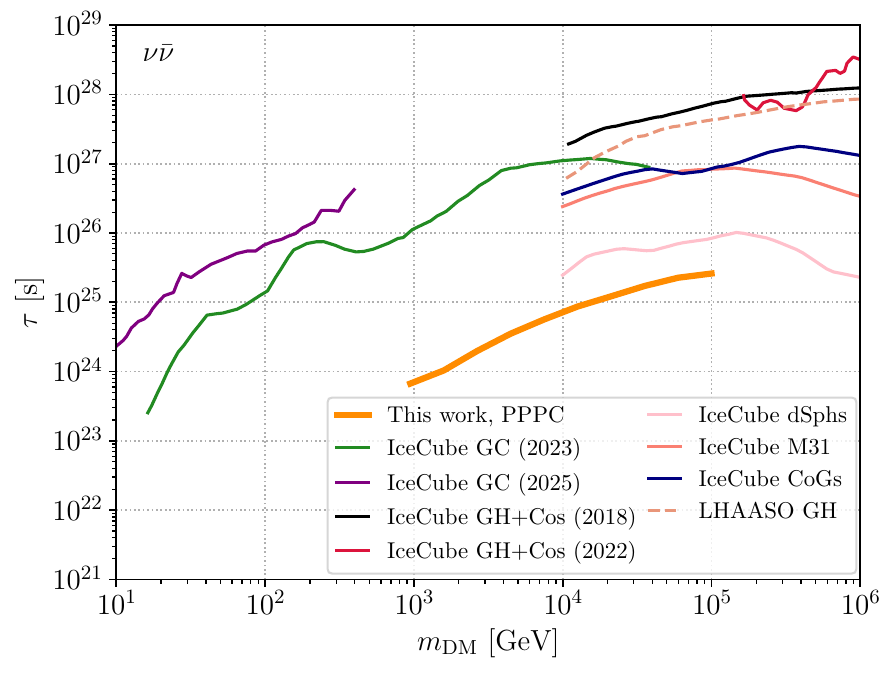}
    \includegraphics[width=0.46\textwidth]{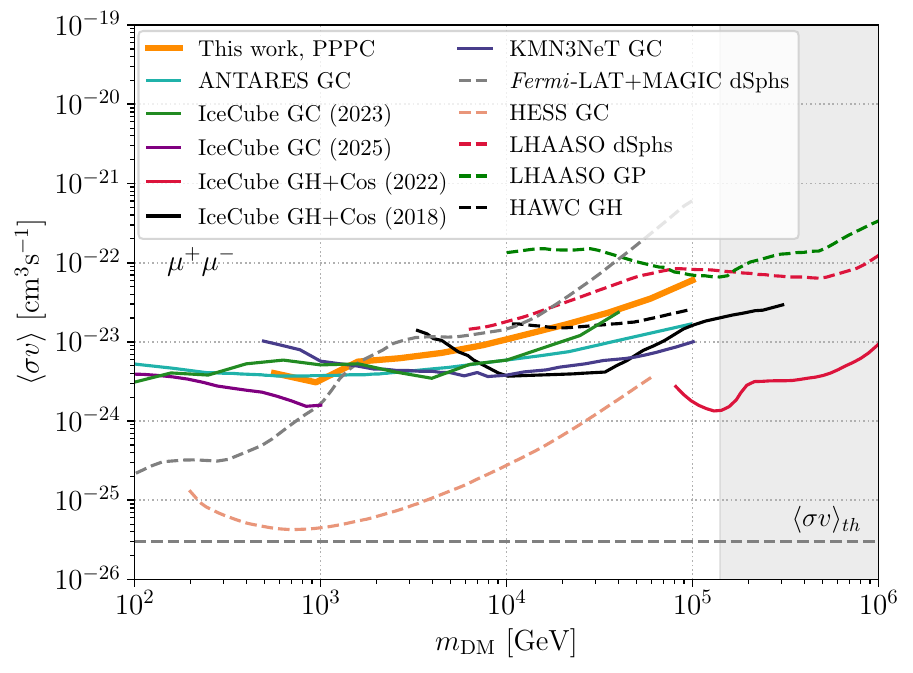}
    \includegraphics[width=0.46\textwidth]{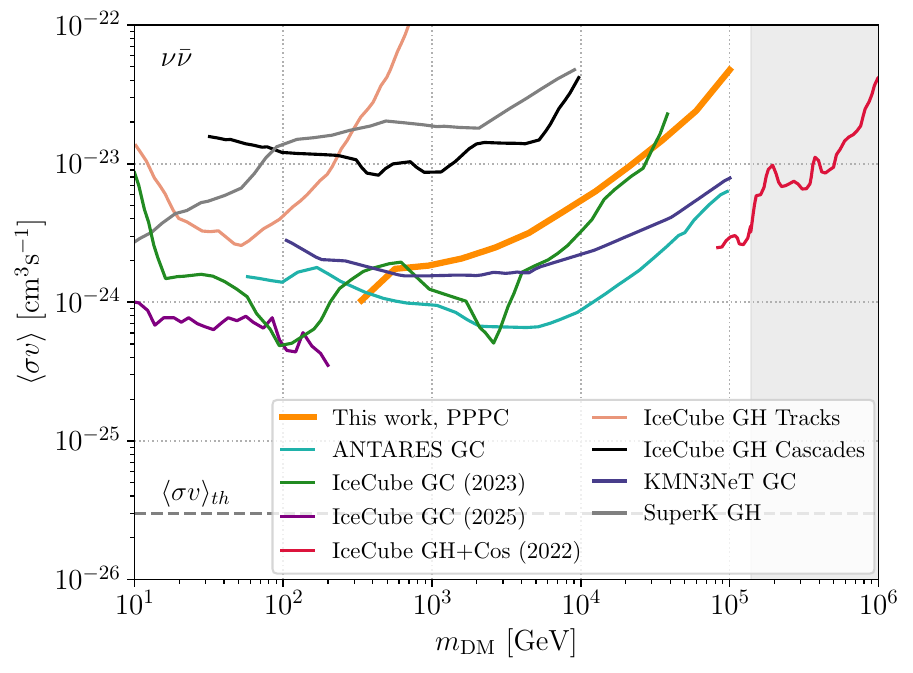}
    \caption{\footnotesize{$90\%$ C.L. (yellow thick line) lower limits on the decay lifetime $\tau$ (top row) and upper limits on the annihilation cross-section $\langle \sigma v \rangle$ (bottom row) for the $\mu$ (left column) and $\nu$ (right column) channels, compared with other neutrino (solid lines) and gamma-ray (dashed lines) data. The constraints correspond to the case of the \texttt{PPPC} spectra computed with the NFW profile with the inclusion of the GDE in the analysis. The grey band in the bottom row indicates when the unitarity limit breaks for thermal cold DM ($m_\mathrm{DM} \gtrsim 140$ TeV). References:  IceCube GC (2023) \cite{2023PhRvD.108j2004A}, IceCube GC (2025) \cite{2025arXiv251100918T}, IceCube GH+Cos (2018) \cite{2018EPJC...78..831A}, IceCube GH+Cos (2022) \cite{2023JCAP...10..003A}, LHAASO GH \cite{2022PhRvL.129z1103C}, \textit{Fermi}-LAT dSphs \cite{2016PhRvD..93j3009B}, HAWC Virgo \cite{2024PhRvD.109d3034A}, HAWC GH \cite{2023JCAP...12..038A}, IceCube dSphs, M31 and cluster of galaxies (CoGs) \cite{2023arXiv230804833J}, ANTARES GC \cite{ALBERT2020135439},  KMN3Net GC \cite{2025JCAP...03..058A}, \textit{Fermi}-LAT+MAGIC \cite{2016JCAP...02..039M}, HESS GC \cite{PhysRevLett.129.111101},  LHAASO dSphs \cite{2024PhRvL.133f1001C}, LHAASO GP \cite{2025arXiv250909609R}, IceCube GH Tracks \cite{2017EPJC...77..627A}, IceCube GH Cascades \cite{2016EPJC...76..531A},  SuperK GH \cite{2020arXiv200505109S}.}}
\label{fig:constraints_comparison} 
\end{figure*}

Finally, the bottom row of Fig.~\ref{fig:constraints_comparison} illustrates how our constraints (yellow thick line) compare with other limits obtained with gamma-ray \cite{2023JCAP...12..038A, 2024PhRvL.133f1001C, 2025arXiv250909609R, 2016JCAP...02..039M, PhysRevLett.129.111101} (dashed lines) and neutrino  \cite{2018EPJC...78..831A, 2023JCAP...10..003A, 2023PhRvD.108j2004A, 2025arXiv251100918T, ALBERT2020135439,  2025JCAP...03..058A, 2020arXiv200505109S, 2017EPJC...77..627A, 2016EPJC...76..531A} (solid lines) data\footnote{See footnotes \ref{fnote:CL_constraints} and \ref{fnote:nu_channel}.}. The left panel is for the $\mu$ channel and the right panel corresponds to the $\nu$ channel. Comparing our results, we can see that in both channels our results are competitive with the rest of the constraints computed from neutrinos, and in the $\mu$ channel the limits are improved by the gamma-ray limits by HESS \cite{PhysRevLett.129.111101} and \textit{Fermi}-LAT+MAGIC \cite{2016JCAP...02..039M}. We note that, in both KMN3Net \cite{2025JCAP...03..058A} and ANTARES \cite{ALBERT2020135439}, the upper limits are also computed with \texttt{PPPC}, making the previous statement more robust. The grey bands in the Figures indicate when the unitarity limit breaks for thermal cold DM ($m_\mathrm{DM} \gtrsim 140$ TeV). A comparison of our constraints in the $\tau^+\tau^-$ channel also shows that the Ridge constraints are quite competitive with previous IceCube constraints (see \ref{ap:app_tau_constraints_comparison}). Finally, we have noted that after convolving the DM signal with ANTARES energy resolution (Eq.~\ref{eq:energy_dispersion}), the limits are worsened by a factor of a few at most in some channels, particularly for the neutrino channels that can reach $\sim1$ order of magnitude.

\subsection{Sensitivity prospects for KM3NeT to the WIMP scenario}
\label{subsec:uncertainty}

Looking ahead, we investigate what the uncertainty on the current Ridge measurements must be to reach the thermal relic annihilation cross-section $\langle \sigma v \rangle_\mathrm{th} = 3 \times 10^{-26} \ \mathrm{cm}^{3} \mathrm{s}^{-1}$ (or a lifetime of $\tau = 10^{29} \ \mathrm{s}$\footnote{We choose this value because it corresponds to the highest constraints available in the literature.}).
Assuming that the ANTARES best-fit is a realistic model of the neutrino flux, we can try to answer this question. The computation of the uncertainty is defined as the $\%$ of uncertainty needed over the ANTARES best-fit flux so the DM flux is distinguishable from the astrophysical background flux (described by the best-fit power law) expected in the region. With this definition, note that the smaller the uncertainty, the greater the precision required to reach the expected DM flux. Therefore, in cases where there is a clear peak in the DM flux (such as the neutrino channels), this uncertainty could be achievable in future neutrino telescopes. As before, we convolve the DM spectra as Eq.~\ref{eq:energy_dispersion}, with the same selection of $\sigma=2$, assuming an energy resolution of $\sim100\%$. Here, we note that while ANTARES was decommissioned in 2022, KMN3Net is expected to significantly improve this measurement and tighten the constraints once it enters its final array configuration status.

\begin{figure*}[t!] 
    \centering 
    \includegraphics[width=0.46\textwidth]{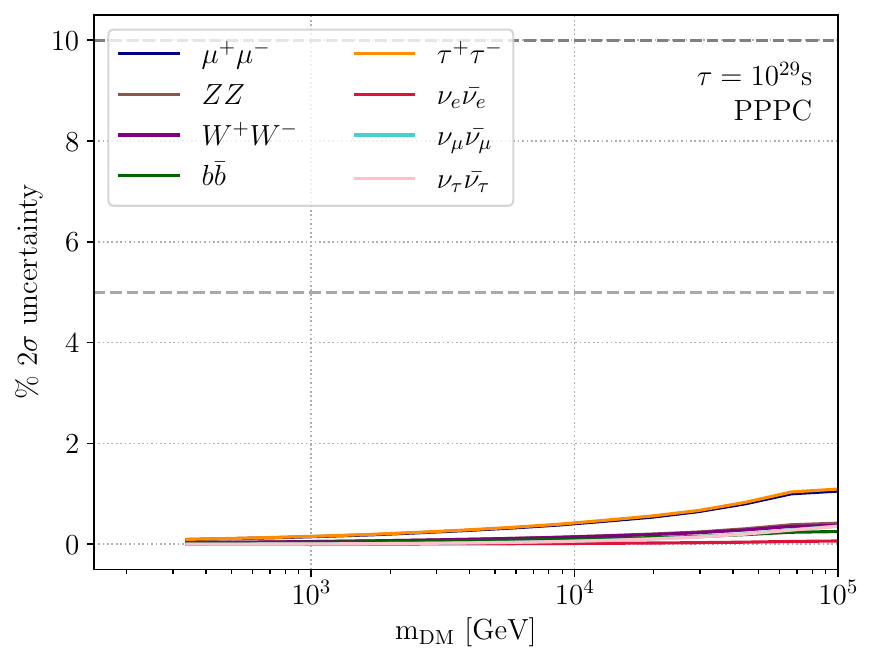}
    \includegraphics[width=0.46\textwidth]{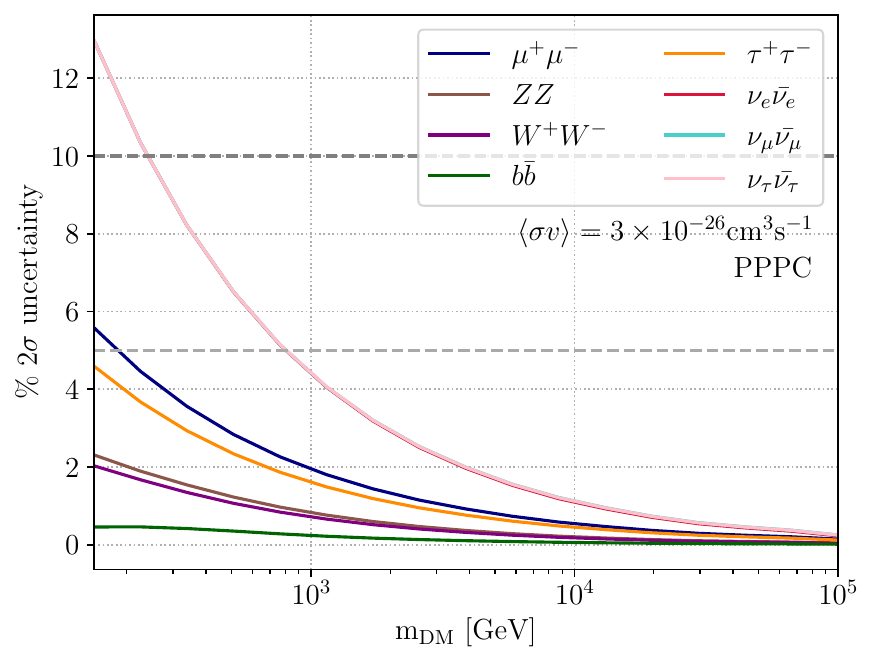}
    \caption{\footnotesize{$2\sigma$ flux uncertainty needed for a future neutrino telescope to be sensitive to a decay lifetime $\tau = 10^{29}$~s (left panel) and the thermal relic cross-section $\langle \sigma v \rangle_\mathrm{th}$ (right panel) for different annihilation channels. The horizontal grey lines represent the expected ideal maximum precision for KMN3Net. The constraints are computed with \texttt{PPPC} and assuming an NFW profile. Given the differences of the flux peaks among the different codes (\texttt{PPPC}, \texttt{CosmiXs}, \texttt{HDMS}), we show the corresponding results in \ref{ap:app_codes_prompt_emission}}.} 
\label{fig:all_uncertainty} 
\end{figure*}

As a benchmark, we take the DM flux computed with the \texttt{PPPC} simulation and with the NFW profile. In Fig.~\ref{fig:all_uncertainty}, we show the uncertainty for all the channels ($\mu$, $b$, $Z$, $\tau$, $W$, and the neutrino channels) for the decaying case (left panel) and annihilating case (right panel). We can see that the uncertainties needed are typically of the order of $\sim5\%$ or less. Interestingly, for the annihilating case in the neutrino channels, the uncertainties grow up to $12\%$ at $m_\mathrm{DM} \sim 300 \ \mathrm{GeV}$. However, this is only true when adopting the \texttt{PPPC} spectra, because in the neutrino channels the peak in the flux is higher than for the other codes. To see a comparison with \texttt{HDMS} and \texttt{CosmiXs}, see \ref{ap:app_codes_prompt_emission}. Note that these percentages are strongly dependent on the instrument’s energy resolution, since the neutrino spectral peaks are significantly smeared. 
Adopting an energy resolution of $\sim 10\%$ would allow the error bars to be larger by a factor of a few, or even 1 order of magnitude in the neutrino channels, to probe the thermal relic cross-section.

In these figures, we indicate a benchmark uncertainty with horizontal grey lines representing the expected ideal maximum precision of KM3NeT, at the level of $\sim5\%$–$10\%$ systematic uncertainty, assuming negligible statistical errors. 
We adopt these values because, even in the unrealistic scenario where statistical uncertainties become subdominant, an irreducible threshold would ultimately limit the precision of these measurements: these are systematic uncertainties predominantly coming from the photomultipliers' quantum efficiency and the photon absorption length in the Mediterranean waters~\cite{2025icrc.confE1039F}. Taking the systematic uncertainties that have been estimated for KM3NeT/ARCA21 in Ref.~\cite{2025icrc.confE1039F}, one finds that the thermal relic cross-sections will never be within the reach of KM3NeT through the Ridge observations. This happens for all channels except for the neutrino channels in masses between $\sim300$~GeV and $\sim10$~TeV, although this result is highly dependent on the overall astrophysical emission of the region and the assumed DM density profile. 
The fact that, even under the most optimistic assumptions and smallest uncertainties, we cannot probe the thermal relic cross-section for most channels suggests that Galactic neutrino observations are unlikely to reach this sensitivity in the foreseeable future.

\section{Probing other interesting DM candidates with the  Ridge observations}
\label{sec:OtherDM}

In this section, we discuss other cases where TeV neutrino observations from the Ridge can be relevant besides the WIMP vanilla case.  First, we explore the potential of the Ridge observations to probe branon WIMP DM, for the first time in the context of neutrino measurements. Then, we focus on a few examples of concrete particle models that produce a viable heavy sterile neutrino as a DM candidate and stress that these observations constitute an optimal tool for constraining the decay of these particles. 

\subsection{Constraints on branon DM}
\label{subsec:Branons}

Extra-dimensional \textit{brane-world} models may produce thermal cold DM candidates with masses up to 100 TeV. In these models, while the SM particles live in a three-dimensional hypersurface (brane) embedded in the higher-dimensional space, gravitons are also able to propagate in the different $N$ extra dimensions, such that the total bulk space has $D = 4 + N$ dimensions. The scale of gravity is then given by $M_D$, rather than the Planck scale, and can range from the electroweak scale ($\sim250$ GeV) to even larger than the TeV scale, which makes them interesting targets for current neutrino experiments like IceCube and KM3NeT. In this scenario, the fluctuations of these branes create massive particles, the branons, which naturally can account for the DM content of the Universe \cite{2003PhRvL..90x1301C, 2003PhRvD..68j3505C, 2006PhRvD..73c5008C}.

\begin{figure*}[t!] 
    \centering 
    \includegraphics[width=0.46\textwidth]{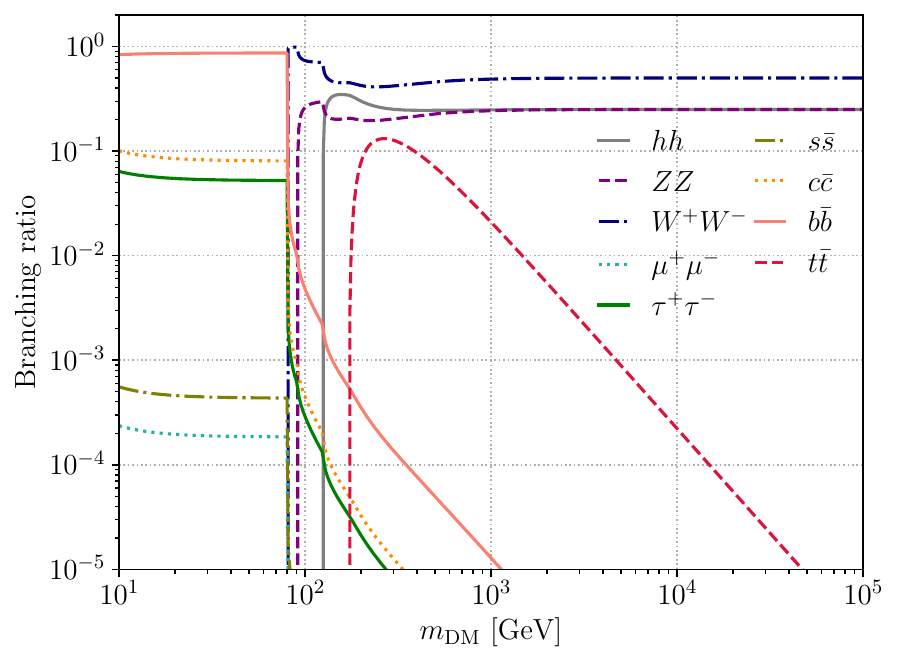}
    \includegraphics[width=0.46\textwidth]{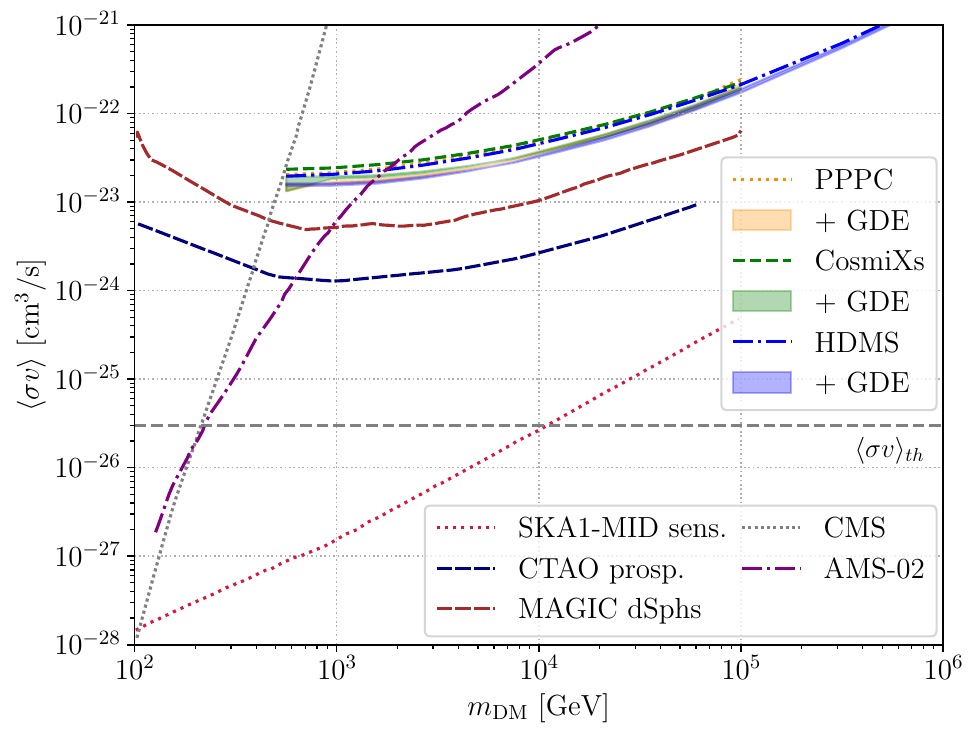}
    \caption{\footnotesize{Left panel: branon annihilation branching ratios into different channels. Note that, for each DM mass $m_\mathrm{DM}$, the branching ratios sum up to 1. Right panel: similar to Fig. \ref{fig:constraints_codes}, we show the upper limits to the branon annihilation cross-section parameter $\langle \sigma v \rangle$, assuming the NFW DM density profile and we compare it with other experiments, such as the projected upper limits with CTAO observing the dwarf Spheroidal galaxy Draco \cite{2020JCAP...10..041A} (dark blue dot-dashed line), MAGIC's dSphs survey (brown dot-dashed line) \cite{Abe_2025}, the SKA1-MID projected detectability region on Draco \cite{2020PDU....2700448C} (dotted red line), CMS (dotted grey line) \cite{2014arXiv1410.8812C} and with AMS-02 data \cite{Cembranos:2017eie} (purple dashed line).}}
\label{fig:branons} 
\end{figure*}

These candidates would annihilate, producing emissions like other WIMPs generate, but with branching ratios $\mathrm{BR}_i$ that depend on the DM mass and the characteristics of the branes (e.g., their tension $f$) \cite{2003PhRvL..90x1301C, 2003PhRvD..68j3505C, 2006PhRvD..73c5008C}. Following \cite{2003PhRvD..68j3505C}, at leading order in $m_\mathrm{DM}/T$, the thermally averaged annihilation cross-section can be expressed as  $\langle \sigma v \rangle\simeq f^8 \sum_i d_{i}$, where $i$ correspond to the particular annihilation channel (fermions $\psi$, massive gauge fields $G$ and complex scalar fields $\Phi$) and $d_i$ are given by:
\begin{equation}
\begin{aligned}
    d_{\psi} = \frac{m_\mathrm{DM}^2 m_\psi^2\left(m_\mathrm{DM}^2-m_\psi^2\right)}{16 \pi^2}  \sqrt{1-\frac{m_\psi^2}{m_\mathrm{DM}^2}} \\  
    d_{G}=\frac{m_\mathrm{DM}^2 \left(4 m_\mathrm{DM}^4-4 m_\mathrm{DM}^2 m_G^2+3 m_G^4\right)}{64 \pi^2} \sqrt{1-\frac{m_G^2}{m_\mathrm{DM}^2}} \\
    d_\Phi = \frac{m_\mathrm{DM}^2 \left(2 m_\mathrm{DM}^2-m_{\Phi}^2\right)^2}{32 \pi^2} \sqrt{1-\frac{m_{\Phi}^2}{m_\mathrm{DM}^2}}
\end{aligned}
\end{equation}

These heavy DM candidates have been searched for by different high-energy gamma-ray experiments, and have received special attention from very-high-energy observatories, such as the next-generation Cherenkov Telescope Array Observatory (CTAO) \cite{2020JCAP...10..041A} and MAGIC \cite{Abe_2025}. However, they have not been explored by means of neutrino observations. Therefore, we also explore the Ridge measurements to constrain DM composed of the branons. We show in the left panel of Fig.~\ref{fig:branons} the branon branching ratios as a function of the DM mass and, in the right panel, the cross-section $\langle \sigma v \rangle$ $90\%$ C.L. upper limits computed with the different spectra codes (yellow for \texttt{PPPC}, green \texttt{CosmiXs} and blue \texttt{HDMS}) and with the inclusion of GDE (colored bands) or with the DM only case (solid lines). We also show the comparison of our constraints with other experiments, such as the projected upper limits with CTAO observing the dwarf Spheroidal galaxy Draco \cite{2020JCAP...10..041A} (dark blue dot-dashed line), MAGIC's dSphs survey (brown dot-dashed line) \cite{Abe_2025}, the SKA1-MID projected detectability region on Draco \cite{2020PDU....2700448C} (dotted red line), CMS (dotted grey line) \cite{2014arXiv1410.8812C} and with AMS-02 \cite{Cembranos:2017eie} data (purple dashed line). As  \texttt{HDMS} yields the most extended results in mass, reaching $m_\mathrm{DM}>100$ TeV, and with very little difference from the other codes, we take the \texttt{HDMS} code as our benchmark for the branons results. Also, the Figure illustrates the potential of neutrino measurements to constrain branon DM, as the ANTARES results are competitive with the current astrophysical constraints set by MAGIC. It can be noted that the best constraints come from the SKA1-MID projected detectability region \cite{2020PDU....2700448C}, computed from the synchrotron radio emission of the secondary particles created by the branon annihilation. However, the results assume 1000h of observation and are subject to important uncertainties: diffusion of the charged particles in the dwarf Spheroidal galaxy Draco and the modeling of its magnetic field, among others.

\begin{figure}[t!] 
    \centering 
    \includegraphics[width=0.46\textwidth]{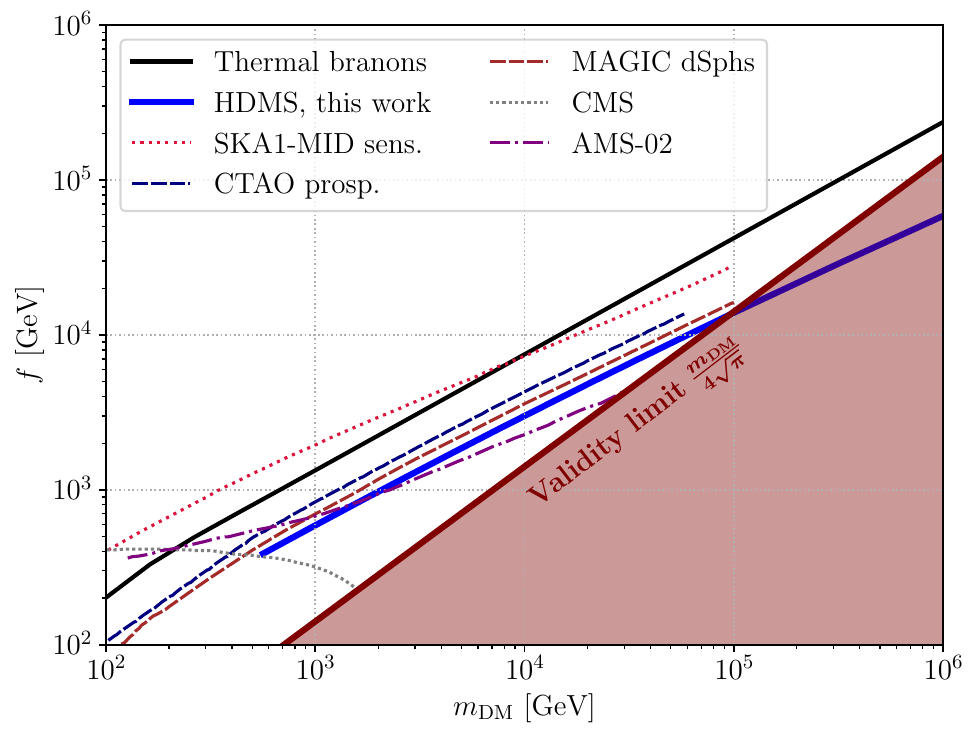}
    \caption{\footnotesize{Lower limits to the branon tension $f$, where the black solid line represents the precise $f$ value for branons to be the entire DM content. With the same color scheme as the right panel of Fig. \ref{fig:branons}, we show other constraints from the literature. With the brown region, we illustrate the validity limit of the theory used \cite{2006PhRvD..73c5008C}.}}
\label{fig:branons_tension} 
\end{figure}

Finally, as the branon annihilation rate is directly dependent on the brane tension $f$, we can translate the constraints obtained on $\langle \sigma v\rangle$ to lower limits in $f$. We show in Fig.~\ref{fig:branons_tension} our final constraints (blue solid line, corresponding to \texttt{HDMS}) compared with the literature, similar to the right panel of Fig.~\ref{fig:branons}. Furthermore, assuming that DM is only made of branons, then it can be computed the exact value of $f$ to yield the thermal relic density $\Omega_\mathrm{DM}$ (represented by the black solid line).  Finally, the thick brown line shows the validity limit of the weakly coupled limit of the theory $f \gtrsim m_\mathrm{DM}/(4 \sqrt{\pi})$, where the approximations to estimate the branching ratios break \cite{2006PhRvD..73c5008C}. As before, our results are competitive with the current MAGIC constraints, allowing us to rule out a big portion of the parameter space. We note that the energy resolution is critical in this case, since branon annihilation at the energies that we are exploring is dominated by W$^{\pm}$ channels, whose corresponding flux is significantly affected when accounting for the energy resolution kernel. Improving the energy resolution from $\sigma = 2$ to values of $\sigma = 1.1$, these constraints reach the limits set by MAGIC.

\subsection{Very heavy sterile neutrinos}
\label{subsec:Snu}

\begin{figure*}[t!] 
    \centering 
    \includegraphics[width=0.46\textwidth]{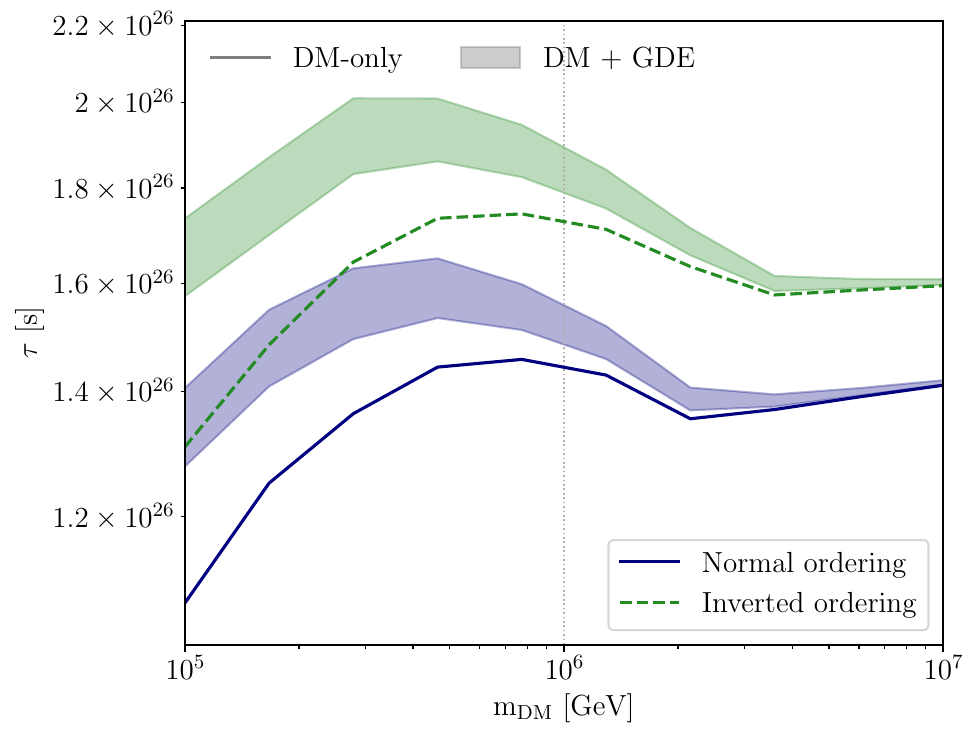}
    \includegraphics[width=0.46\textwidth]{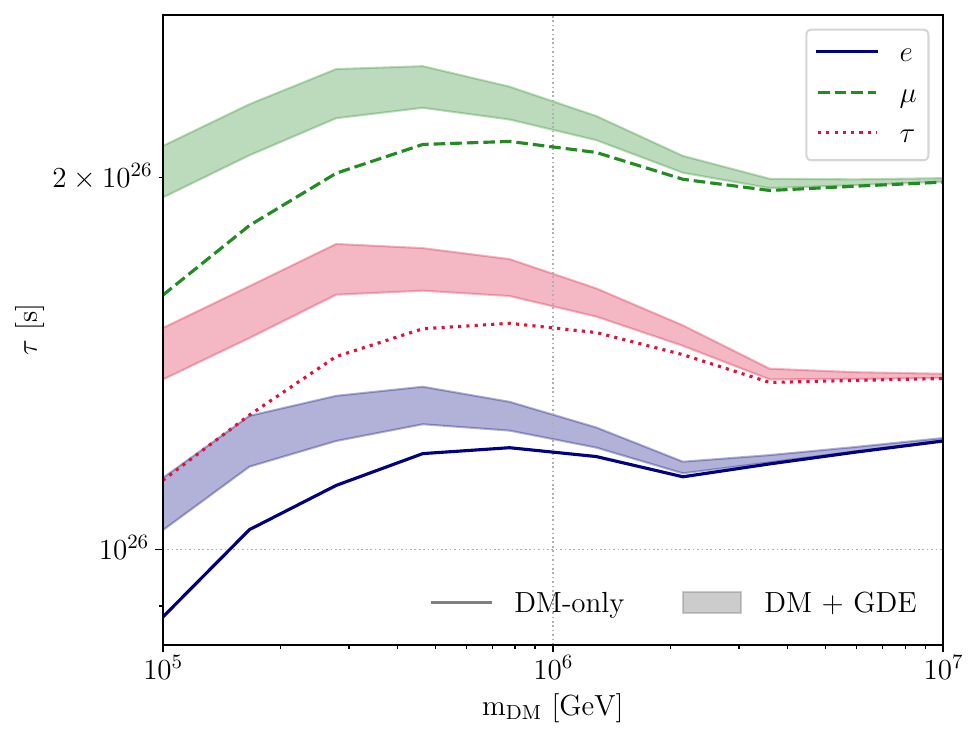}
    \caption{\footnotesize{Lower limits ($90\%$ C.L.) for the decay lifetime of sterile neutrino DM for the total model (left panel: blue normal ordering, green inverted) and separated into the three different lepton channels (right panel: electron $e$ in blue, muon $\mu$ in green and tau $\tau$ in red). The colored bands represent the constraints computed with the inclusion of the GDE, and the lines represent the DM-Only spectra. These constraints are computed assuming \texttt{HDMS} and the NFW profile. For the constraints on the individual neutrino channels and the Yukawa couplings, see \ref{ap:app_ind_constraints_neutrinoful}.}}
\label{fig:neutrinoful} 
\end{figure*}

In a general framework, several models of heavy sterile neutrinos constituting the DM abundance today are expected to oscillate or decay, generating a continuous emission of active neutrinos. One of the most popular frameworks predicting the existence of heavy sterile neutrinos is the $\mathrm{B-L}$ extension of the SM. Inspired by the model presented in Ref.~\cite{2014JHEP...07..044H}, we explore the possibility of DM constituted by a heavy right-handed sterile neutrino created from a $U(1)_\mathrm{B-L}$ extended SM. In this scenario, the sterile neutrinos are produced from the $\mathrm{B-L}$ Higgs field that, after spontaneously breaking the $U(1)_\mathrm{B-L}$ symmetry, oscillates in the minimum of the potential giving masses to the right-handed neutrinos through Yukawa couplings. This coupling, at the same time, allows the inflaton to decay into the right-handed neutrinos, providing a natural production of such species. Besides, the baryon asymmetry can be explained by the decay of these right-handed neutrinos, while the lightest right-handed neutrino emerges as a DM candidate. Interestingly, if the reheating process in the early Universe is dominated by the decay of the inflaton into the second lightest right-handed neutrinos, the lightest right-handed neutrino —and, thus, also the DM candidate— gains a mass of the order of PeV \cite{2014JHEP...07..044H}. In this context, heavy sterile neutrinos represent optimal targets for neutrino DM searches.

Within this framework, this DM candidate is coupled to the SM through Yukawa couplings $y_\nu^l$ (where $l$ indicates the lepton family; $l$= e, $\mu$, $\tau$), whose main decay mode would be into a lepton and a $W$ boson, or a neutrino and a $Z$/$h$ boson. These couplings are directly proportional to the DM decay rates $\Gamma$ as \cite{2014JHEP...07..044H}:
\begin{equation}
\begin{aligned}
    \Gamma(\nu_s \rightarrow l^{\pm} W^{\mp}) = \frac{|y_\nu^l|^2 m_{\mathrm{DM}}}{16\pi} \left(1 - \frac{m_{W}^2}{m_{\mathrm{DM}}^2} \right)^2 \left(1 + \frac{2m_{W}^2}{m_{\mathrm{DM}}^2} \right)  \\ 
    \Gamma(\nu_s \rightarrow \nu_l(\bar{\nu}_l)Z) = \frac{|y_\nu^l|^2 m_{\mathrm{DM}}}{32\pi} \left(1 - \frac{m_{Z}^2}{m_{\mathrm{DM}}^2} \right)^2 \left(1 + \frac{2m_{Z}^2}{m_{\mathrm{DM}}^2} \right) \\
    \Gamma(\nu_s \rightarrow \nu_l(\bar{\nu}_l)h) = \frac{|y_\nu^l|^2 m_{\mathrm{DM}}}{32\pi} \left(1 - \frac{m_{h}^2}{m_{\mathrm{DM}}^2} \right)^2,
\end{aligned}
\label{eq:rates_neutrinoful}
\end{equation}
\noindent where $\Gamma_\mathrm{total}$ is the inverse of the lifetime $\tau$ of the particle. These decay channels can be strongly constrained by the Ridge observations, whose limits can be expressed in the form of their Yukawa couplings $y_\nu^l$, or lifetime $\tau$, as a function of the sterile neutrino mass $m_\mathrm{DM}$, that we take of the order of the PeV \cite{2014JHEP...07..044H}. We also note that similar channels can also be the main decay channels for other kinds of heavy scalar DM particles, which makes neutrino telescopes exceptionally interesting to study such DM candidates.

Considering very high DM masses, in the limit of $m_\mathrm{DM} \gg m_W, m_Z, m_h$  in Eq.~\ref{eq:rates_neutrinoful}, the boson mass dependence disappears, and therefore the Yukawa couplings can be directly related to the decay rate. Following the same logic, considering all the decay rates of a single lepton channel $l = e, \mu, \tau$, the decay rate is approximated by:
\begin{equation}
    \Gamma_{{\nu}_s}^l = \frac{m_{\mathrm{DM}}}{4 \pi} |y_{\nu}^l|^2 = 10^{-28} \mathrm{s^{-1}} \left(\frac{|y_\nu^l|}{10^{-29}}\right)^2 \left(\frac{m_{\mathrm{DM}}}{8\  \text{PeV}}\right)
    \label{eq:gamma_to_yukawa}
\end{equation}

Note that the complete model is given by the total decay rate, computed as a summation over the three lepton channels. These branching ratios depend on the mass hierarchy ordering of the neutrinos (normal or inverted), which we list as follows~\cite{2014JHEP...07..044H}:
\begin{equation}
\begin{aligned}
    \text{Normal ordering:}\quad 
    (\mathrm{BR}(e), \mathrm{BR}(\mu), \mathrm{BR}(\tau))
    &= \\
    \left(0.68,\ 0.24 + 0.02\cos\delta,\ 0.08 - 0.02\cos\delta\right) \\
    \text{Inverted ordering:}\quad
    (\mathrm{BR}(e), \mathrm{BR}(\mu), \mathrm{BR}(\tau))
    &= \\
    \left(0.02,\ 0.38,\ 0.60\right),
\end{aligned}
\label{eq:neutrinoful_lepton_BR}
\end{equation}
\noindent where $\delta$ is the CP-violating phase, which we set to 0 following~\cite{2014JHEP...07..044H}. Once Eq.~\ref{eq:neutrinoful_lepton_BR} is taken into account in the decay flux, all three Yukawa couplings are uniquely related and, similarly to Eq.~\ref{eq:gamma_to_yukawa}, the total decay rate can also be expresses as $\Gamma_{\nu_s} = \frac{m_{\mathrm{DM}}}{4 \pi} |y_{\nu}|^2$.

As we are in the very heavy DM masses, we make use of \texttt{HDMS} to compute the annihilation neutrino spectra, and we show in the left panel of Fig.~\ref{fig:neutrinoful} the lower limits on the decay lifetime for the full model (blue normal ordering and green inverted), and also with the inclusion of the GDE (colored bands) and DM-Only constraints (lines). In the right panel, we show the same results but considering only each leptonic channel individually (electron $e$ in blue, muon $\mu$ in green and tau $\tau$ in red). Finally, for the limits for each individual channel given by Eq.~\ref{eq:rates_neutrinoful} — i.e., assuming only one decay channel instead of the full combination— and the constraints on the specific Yukawa couplings, see \ref{ap:app_ind_constraints_neutrinoful}. We note that a dedicated study with IceCube or KM3NeT observations with better energy resolution and accounting for backgrounds simultaneously would likely improve these constraints.

Other extensions of the SM featuring an L$_{\mu}$-L$_{\tau}$ model have received much attention recently. These models predict the existence of a massive gauge vector boson, $Z'$, that would exclusively couple to leptons of the second and third families. An interesting way to constrain the models that produce this Z' boson is by looking at the spectral distortions that such bosons can produce in the neutrino emission via resonant interactions at the Z' mass~\cite{Zprime2}. Although this has been done using IceCube data, it has never been explored in the context of ANTARES data from the Ridge. This is left for a future project. We note that similar heavy bosons can appear naturally to explain the neutrino masses from the sea-saw mechanism~\cite{Roland_2015, Roland:2015yoa}, which makes neutrino observations also interesting for the more popular case of keV sterile neutrinos.

We finally note that, besides models where a heavy sterile neutrino constitutes the DM, one of the most natural scenarios where $\sim$TeV neutrino observations can be relevant is for variations of extra-dimensional Kaluza-Klein (KK) DM models, where the KK neutrino is the lightest KK particle~\cite{Servant_2003}. This makes clear that TeV-PeV observations from the GC can be very important to constrain a variety of particle physics scenarios.

\section{Conclusions}
\label{sec:conclusions}

In this work, we have explored the potential of diffuse neutrino observations from the Galactic Ridge as a probe of particle DM. Using the latest ANTARES measurements of the Ridge neutrino flux, defined in the region $|l| < 30^\circ$, $|b| < 2^\circ$, we have performed a systematic study of DM signals in this region, including both annihilating and decaying scenarios, and compared them against the expected astrophysical GDE. Although the statistical significance of the current ANTARES detection is modest \cite{2023PhLB..84137951A}, the availability of flux measurements and conservative upper limits allows one to derive meaningful and competitive constraints on a wide range of DM models.
A central result of this work is that Galactic neutrino observations represent a powerful and still underexplored channel for indirect dark matter searches, with the potential to improve upon existing neutrino constraints. We further demonstrate that this region of the sky can provide complementary and, in some cases, competitive bounds from neutrino data. 

We have considered a broad set of DM scenarios in which neutrino observations play a central role. Beyond standard WIMP annihilation and decay channels, we have explored other models where neutrinos are among the dominant or most distinctive final states, including branon WIMP DM and very heavy sterile neutrinos. In our analysis, we systematically compare the expected results from popular state-of-the-art cross-sections for the neutrino prompt-emission, as well as consider different DM density profiles, from cuspy with realistic realizations of DM spikes to cored profiles. While uncertainties in the Ridge measurements remain significant, we have shown that these observations lead to interesting constraints across a wide range of DM masses, from 100 GeV to 1 PeV. Moreover, the inclusion of a physically motivated GDE model does not alter our conclusions significantly. We have shown that the limited energy resolution of the ANTARES neutrino measurements has a significant impact on channels of direct neutrino production, although the resulting limits remain comparable to previous results.
Moreover, the Ridge constraints are quite competitive with current IceCube constraints for WIMP DM annihilating into the $\tau^{\pm}$, $\mu^{\pm}$ and $\nu\bar{\nu}$ channels. Meanwhile, in the branon case, the constraints presented in this work are contending with state-of-the-art constraints obtained from gamma-ray telescopes. In the more general WIMP annihilating case, our constraints yield results similar to other neutrino experiments, stressing the potential of the neutrino measurements of the Galactic Ridge.

We emphasize that these constraints are derived from ANTARES measurements estimated under the assumption of a power-law–dominated flux. This approach is feasible because the flux upper limits are determined independently in each energy bin. Nevertheless, explicitly incorporating a DM contribution within the ANTARES template analysis would likely lead to more stringent constraints. We expect that a combined analysis of raw data, incorporating both standard astrophysical neutrino emission and these potential DM signals, would improve these limits by an order of magnitude, surpassing those derived from Galactic halo analyses using the same approach. In this regard, template-fit procedures can significantly reduce uncertainties and become leading constraints from neutrinos.

Looking ahead, the results presented here should be regarded as conservative benchmarks. Significant improvements are expected from current and future neutrino telescopes. Dedicated Galactic Ridge analyses by IceCube, benefiting from increased exposure and improved reconstruction techniques, will already strengthen the constraints derived from existing data. Even more promising are the prospects offered by KM3NeT/ARCA, whose location and detector design are optimized for observations of the Southern sky and the GC region. In the longer term, IceCube-Gen2 will further enhance sensitivity at multi-TeV and PeV energies, allowing Ridge measurements to probe deeper into DM parameter space.
A combined analysis of ANTARES and the current KM3NeT dataset is expected to yield a statistical improvement of up to 30–40\%. Meanwhile, the systematic uncertainties might be reduced to a minimum level of $\sim5$\% in future ARCA observations.

In summary, this work highlights the Galactic Ridge as a promising laboratory for indirect DM searches with neutrinos and emphasizes the use of Galactic neutrino measurements to improve bounds in neutrino channels. By exploiting existing ANTARES measurements and extending the analysis to a wide class of particle physics models, we have demonstrated both the current constraining power and the substantial future potential of Ridge observations. As neutrino telescopes continue to improve in sensitivity and control of systematics, this approach is expected to play a key role in testing DM scenarios at energies and interaction strengths that are otherwise inaccessible.

\section*{Acknowledgements}

The authors want to thank Francesco Filippini and all the DAMASCO group\footnote{\href{https://projects.ift.uam-csic.es/damasco/}{https://projects.ift.uam-csic.es/damasco/}} for fruitful discussions about the Galactic ridge measurements. The work of JZP and PDL is supported by the grants PID2021-125331NB-I00 and CEX2020-001007-S, funded by MCIN\slash AEI (10.13039\slash501100011033), and by ``ERDF A way of making Europe'', and the MULTIDARK Project RED2022-134411-T. JZP's contribution to this work has been supported by \textit{FPI Severo Ochoa} PRE2021-099137 grant and the PID2024-155874NB-C21 grant. This work is also supported by the project PID2022-139841NB-I00 funded by MICIU\slash AEI\slash10.13039\slash501100011033 and by ERDF/EU.
PDL has been supported by the Juan de la Cierva JDC2022-048916-I grant, funded by MCIU\slash AEI\slash10.13039\slash501100011033 European Union "NextGenerationEU"/PRTR, and is currently supported by Ramón y Cajal RYC2024-048445-I grant, which is funded by MCIU\slash AEI\slash10.13039\slash501100011033 and FSE+.

\appendix

\section{Energy resolution discussion}
\label{ap:energy_resolution}

\begin{figure*}[t!]
    \centering 
    \includegraphics[width=0.9\textwidth]{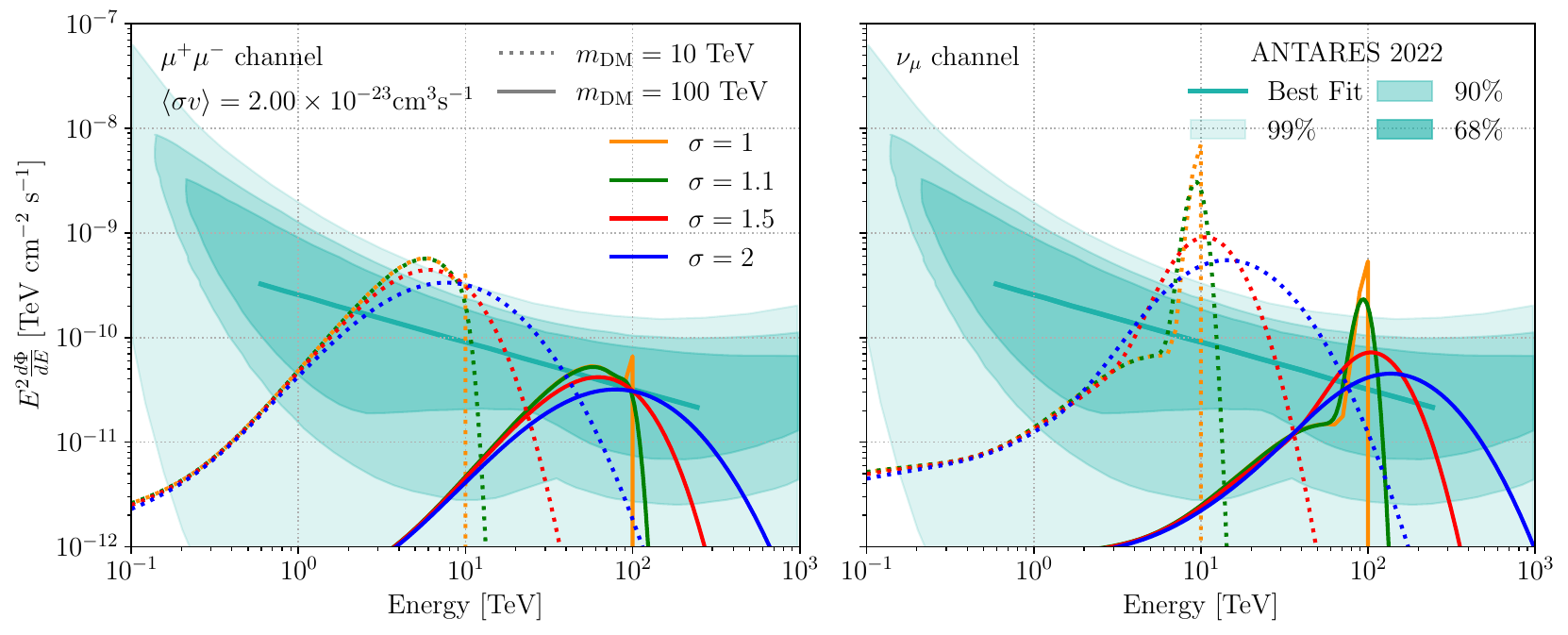}
    \caption{\footnotesize{ANTARES 2022 neutrino flux (light blue bands) compared with the expected neutrino flux in the Galactic Ridge from DM annihilation to the $\mu^+\mu^-$ (left panel) and $\nu_\mu$ channel (right panel). We show the spectra for two different masses ($m_\mathrm{DM} = 10$ TeV (dotted lines) and 100 TeV (solid lines)) and computed with four different energy resolutions: $\sigma = 1$—i.e., perfect energy resolution— (yellow lines), $\sigma =1.1$ (green lines), $\sigma =1.5$ (red lines), and the benchmark $\sigma = 2$ (blue lines). The DM spectra assume an NFW profile in both panels, computed with \texttt{PPPC}.}}
\label{fig:spectra_sigmas} 
\end{figure*}

The dependence of the observed neutrino flux on the energy resolution is especially significant when studying spectra with high peaks, such as those from annihilation neutrino channels. Following the convolution with the energy resolution kernel as in Eq.~\ref{eq:energy_dispersion}, we show in Fig.~\ref{fig:spectra_sigmas} the neutrino fluxes for the $\mu^+\mu^-$ (left panel) and $\nu_\mu$ (right panel) channels and two different masses (10 TeV, dotted lines; 100 TeV, solid lines) varying the $\sigma$ energy resolution parameter. We choose the case of $\sigma = 1$—i.e., perfect energy resolution— (yellow lines), $\sigma =1.1$ (green lines), $\sigma =1.5$ (red lines), and the benchmark $\sigma = 2$ for an energy dispersion of $100\%$ (blue lines). This Figure illustrates how, with an energy resolution of $\sigma=1.1$, most of the original spectra are reproduced with the peak being observable. The spectra shown correspond to the \texttt{PPPC} code, assuming an NFW profile. In general, improving the energy resolution will strengthen the neutrino constraints by a few orders of magnitude across all channels, while improving by $\sim 1$ order of magnitude in the annihilation case and neutrino channels.

\section{DM density profiles}
\label{ap:app_density_profile}
\begin{figure*}[t!] 
    \centering 
    \includegraphics[width=0.9\textwidth]{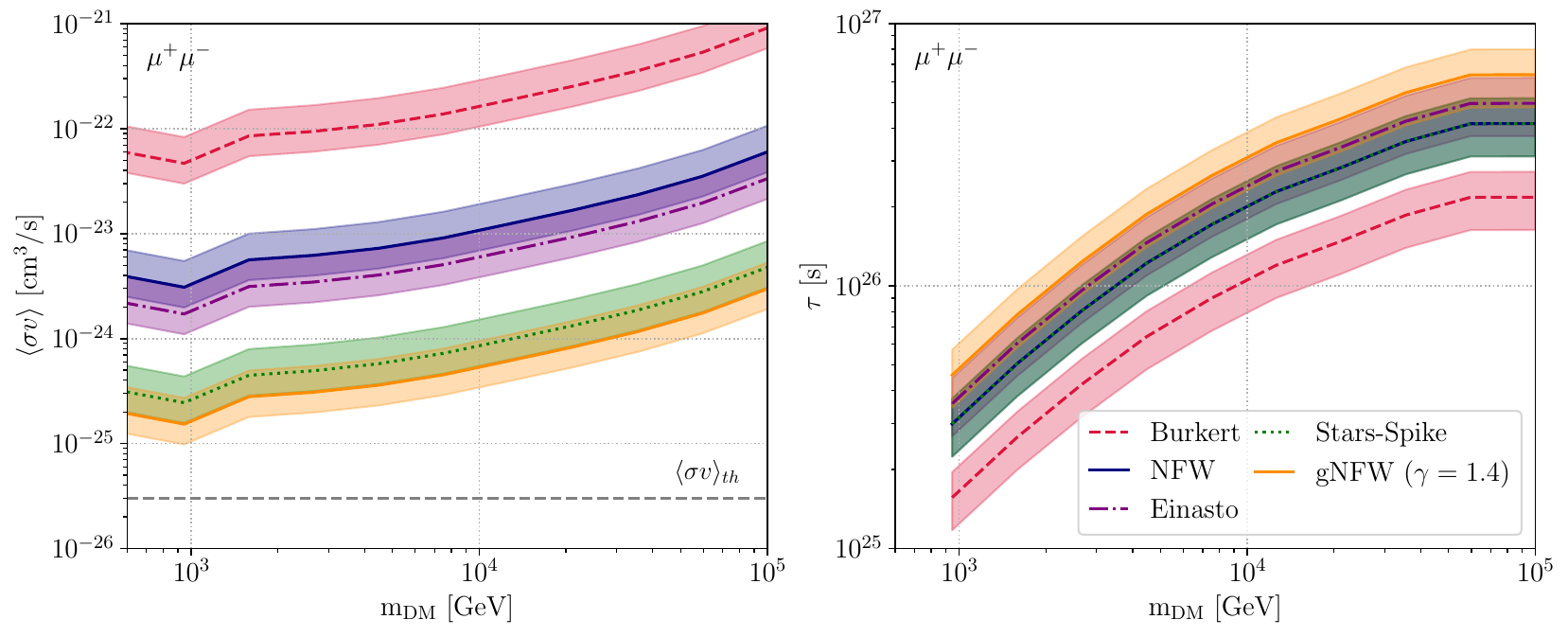}
    \caption{\footnotesize{Dependence of the annihilation (left panel) and decay constraints (right panel) on the Galactic DM density profile assumed for the $\mu$ channel. The blue solid line is for the benchmark NFW profile, the dotted green line is for the Stars-Spike, dashed red Burkert, dot-dashed purple Einasto and yellow gNFW with $\gamma = 1.4$. The colored bands represent the dependence on the DM density profile normalization $\rho_\odot$, which we allow to vary from $0.3$ to $0.5 \ \mathrm{GeV} \, \mathrm{cm^{-3}}$. These constraints are computed with \texttt{PPPC}.}}
    
\label{fig:constraints_other_profiles} 
\end{figure*}
\begin{table*}[t!]
    \begin{center}
        \begin{tabular}{|c|c|c|c|c|c|}
        \hline
        \hline
\textbf{Region}                &  NFW  &  Burkert  & Einasto  & Stars-Spike & gNFW ($\gamma = 1.4$)\\ 
\hline
\hline
\parbox{2.5cm}{ANTARES\\Galactic Ridge} 
& $5.79\times 10^{21}$ & $3.04\times 10^{21}$ & $6.92\times 10^{21}$ & $5.79\times 10^{21}$ & $8.89 \times 10^{21}$ \\
\hline
\parbox{2.5cm}{$\theta_\mathrm{GC} < 2^\circ$}   
& $6.34\times 10^{20}$ & $1.69\times 10^{20}$ & $8.39\times 10^{20}$ & $6.35\times 10^{20}$ & $1.58 \times 10^{21}$ \\
\hline
\parbox{2.5cm}{$l \in [15^\circ,125^\circ]$\\$b \in [-5^\circ,5^\circ]$}   
& $8.34\times 10^{21}$ & $7.95\times 10^{21}$ & $8.50\times 10^{21}$ & $8.34\times 10^{21}$ & $8.35 \times 10^{21}$ \\
\hline
\hline

        \end{tabular}
        \caption{\footnotesize{D-factor values (in $\mathrm{GeV} \, \mathrm{cm}^{-2}$) for the DM density profiles discussed in this work and for different regions.}}
        \label{tab:ROI_Dfactors}
    \end{center}
\end{table*}

\begin{table*}[t!]
    \begin{center}
        \begin{tabular}{|c|c|c|c|c|c|}
        \hline
        \hline
\textbf{Region}                &  NFW  &  Burkert  & Einasto  & Stars-Spike & gNFW ($\gamma = 1.4$) \\ 
\hline
\hline
\parbox{2.5cm}{ANTARES\\Galactic Ridge} 
& $2.30\times 10^{22}$ & $1.52\times 10^{21}$ & $4.14\times 10^{22}$ & $2.90\times 10^{23}$ & $4.64\times 10^{23}$ \\
\hline
\parbox{2.5cm}{$\theta_\mathrm{GC} < 2^\circ$}   
& $1.02\times 10^{22}$ & $9.31\times 10^{19}$ & $1.82\times 10^{22}$ & $2.77\times 10^{23}$  & $4.09\times 10^{23}$ \\
\hline
\parbox{2.5cm}{$l \in [15^\circ,125^\circ]$\\$b \in [-5^\circ,5^\circ]$}   
& $4.15\times 10^{21}$ & $2.59\times 10^{21}$ & $4.88\times 10^{21}$ & $4.15\times 10^{21}$  & $5.68\times 10^{21}$ \\
\hline
\hline

        \end{tabular}
        \caption{\footnotesize{Similar to Tab.~\ref{tab:ROI_Dfactors}, but for the J-factor parameter, units in $\mathrm{GeV}^{2} \, \mathrm{cm}^{-5}$.}}
        \label{tab:ROI_Jfactors}
    \end{center}
\end{table*}
The parameters used to model the DM density profiles are always taken such that the normalization of the profiles yield the local DM density value of $\rho_\odot = 0.4 \ \mathrm{GeV} \, \mathrm{cm^{-3}}$ \cite{2021PDU....3200826B}, with $R_\odot = 8.277 \ \mathrm{kpc}$ \cite{GRAVITY_2021}. For the rest of the parameters, the NFW (Eq.~\ref{eq:NFW_profile}), gNFW (\ref{eq:gNFW_profile}) and Stars-Spike (which is composed of a spike on top of a gNFW profile) are defined with the scale radius $r_\mathrm{s} = 20 \ \mathrm{kpc}$, allowed by the current constraints \cite{2021PDU....3200826B}. In the Einasto profile (Eq.~\ref{eq:Einasto_profile}), we take $\alpha = 0.17$ and $r_\mathrm{s} = 20 \ \mathrm{kpc}$ \cite{Acharyya_2021}. Finally, for the Burkert profile (Eq.~\ref{eq:Burkert_profile}) we take $r_\mathrm{s} = 12.67 \ \mathrm{kpc}$ \cite{2021PDU....3200826B, Cirelli_2011}. With these definitions, we show in Tabs.~\ref{tab:ROI_Dfactors} and \ref{tab:ROI_Jfactors} the D-factor and J-factor values of the DM density profiles defined above for different regions of the Galaxy.

To show the sensibility of the cross-section $\langle \sigma v \rangle$ and lifetime $\tau$ on the DM density profiles used, we show in Fig.~\ref{fig:constraints_other_profiles} the cross-section upper limits (left panel) and lifetime lower limits (right panel) computed with the $\mu^+\mu^-$ channel and \texttt{PPPC} for the different profiles listed above (in blue NFW, green Stars-Spike, red Burkert, purple Einasto and in yellow the gNFW ($\gamma = 1.4$) profile). We also show the uncertainty related to the DM density profile normalization $\rho_\odot$, which we allow to vary from $0.3$ to $0.5 \ \mathrm{GeV} \, \mathrm{cm^{-3}}$. As can be seen, the cuspier the profile, the better the constraints are obtained. This is more evident in the annihilation case, as the J-factor is given by the square of the DM density profile and the constraints can vary up to almost three orders of magnitude. Following Eq.~\ref{eq:Jfactor_general}, all of the results presented in this work can be directly translated to other profiles by rescaling with the correct J-factor/D-factor value.

\section{Constraints with other prompt spectra codes}
\label{ap:app_codes_prompt_emission}

\begin{figure*}[t!] 
    \centering 
    \includegraphics[width=0.46\textwidth]{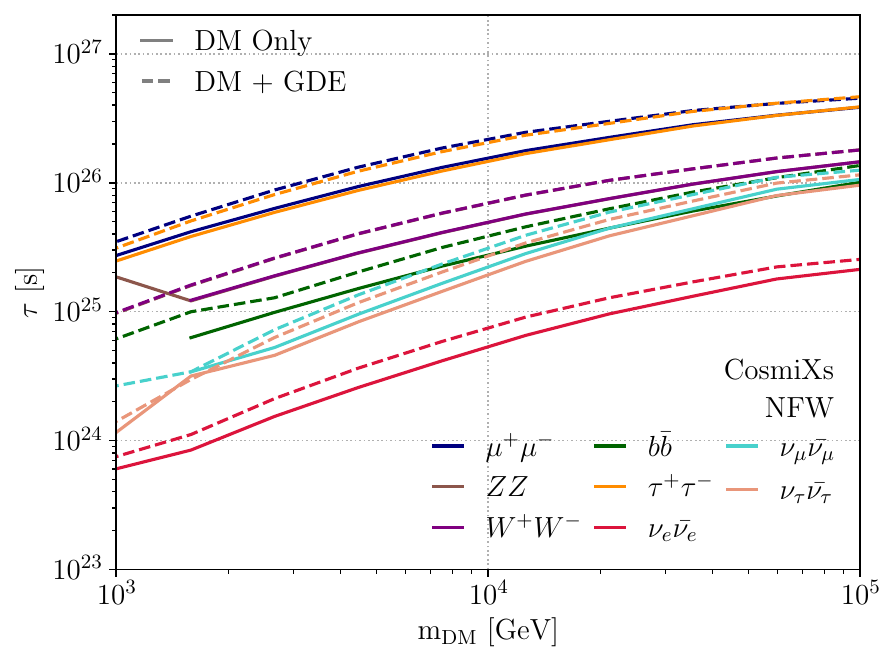}
    \includegraphics[width=0.46\textwidth]{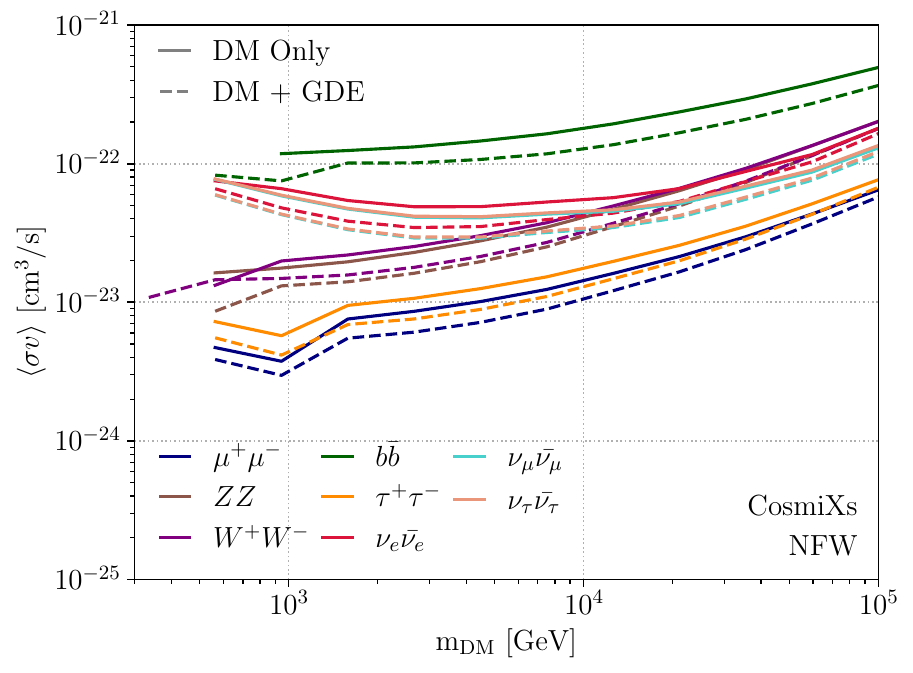}

    \includegraphics[width=0.46\textwidth]{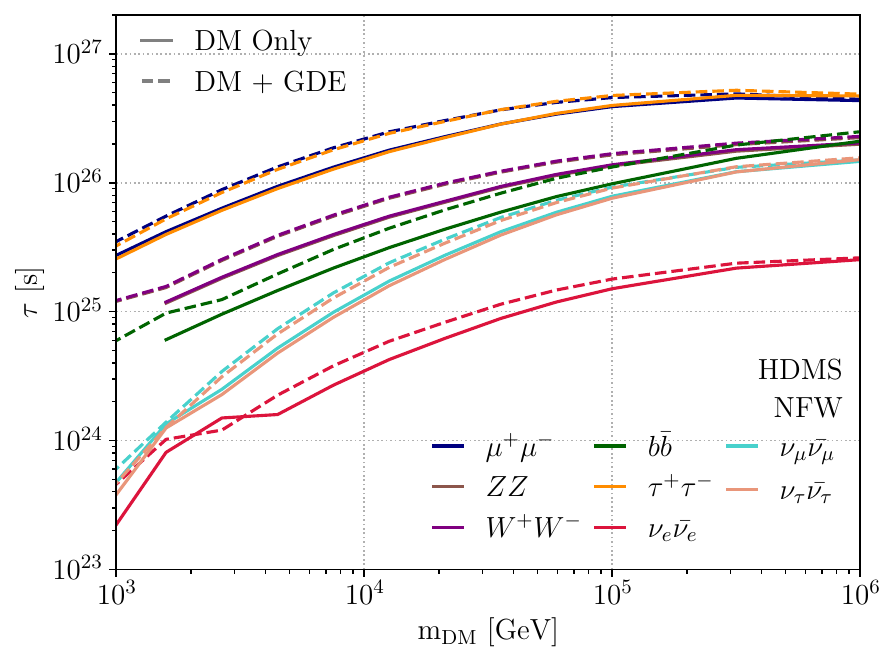}
    \includegraphics[width=0.46\textwidth]{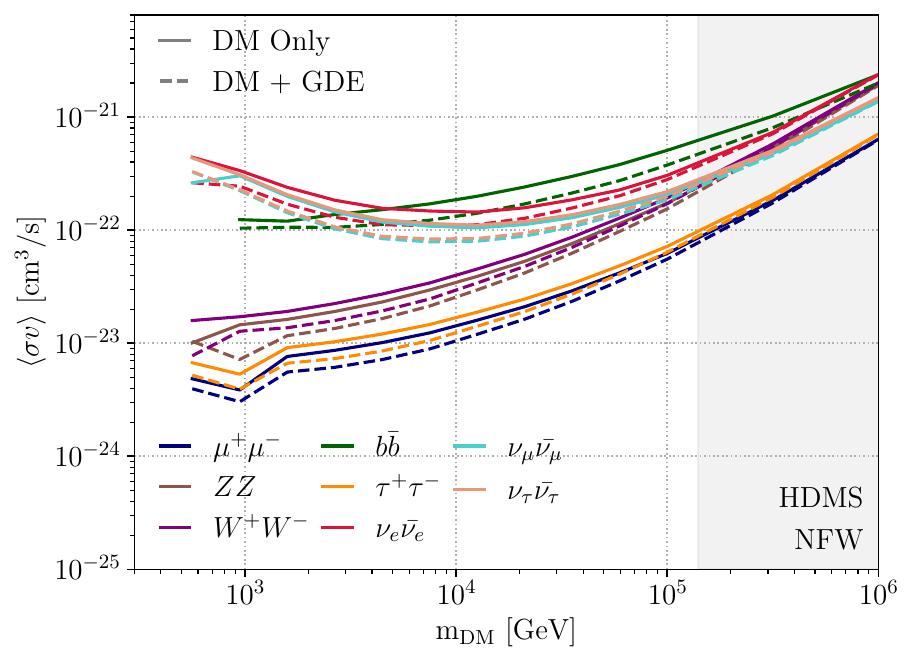}

    \caption{\footnotesize{Same as Fig.~\ref{fig:all_constraints}, but with the spectra computed with \texttt{CosmiXs} (first row) and \texttt{HDMS} (second row). The grey band in the bottom left panel indicates when the unitarity limit breaks for thermal cold DM ($m_\mathrm{DM} \gtrsim 140$ TeV).}}
    
\label{fig:all_constraints_CosmiXs_HDMS} 
\end{figure*}

Similar to Fig.~\ref{fig:all_constraints}, we show the upper/lower limits for the rest of the prompt emission codes in Fig.~\ref{fig:all_constraints_CosmiXs_HDMS}, for the decay case in the left column and annihilation in the right column. In general, the best constraints are obtained with \texttt{PPPC}, with \texttt{CosmiXs} (first row of the figure) and \texttt{HDMS} (second row) yielding similar results except for the neutrino channels, in which \texttt{CosmiXs} gives a brighter $dN/dE$ prompt emission than \texttt{HDMS}, but lower than \texttt{PPPC}. Another difference is that \texttt{HDMS} reaches fluxes with masses greater than $100 \ \mathrm{TeV}$ both for decay and annihilation. To leave a remark on the unitarity limit, we have marked in grey the values of $m_\mathrm{DM}$ that pass the unitarity limit.

\begin{figure*}[t!] 
    \centering 
    \includegraphics[width=0.46\textwidth]{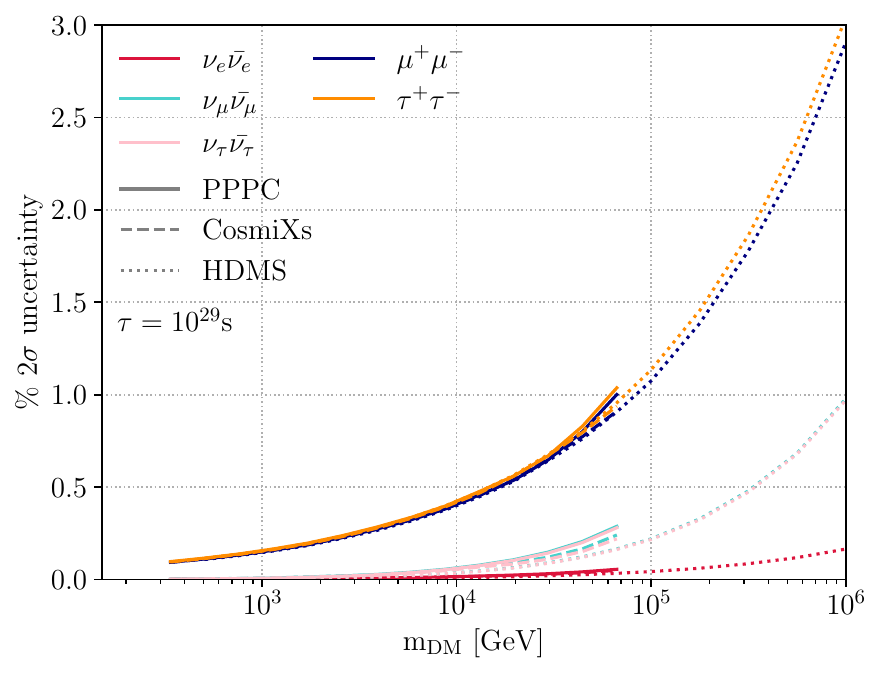}
    \includegraphics[width=0.46\textwidth]{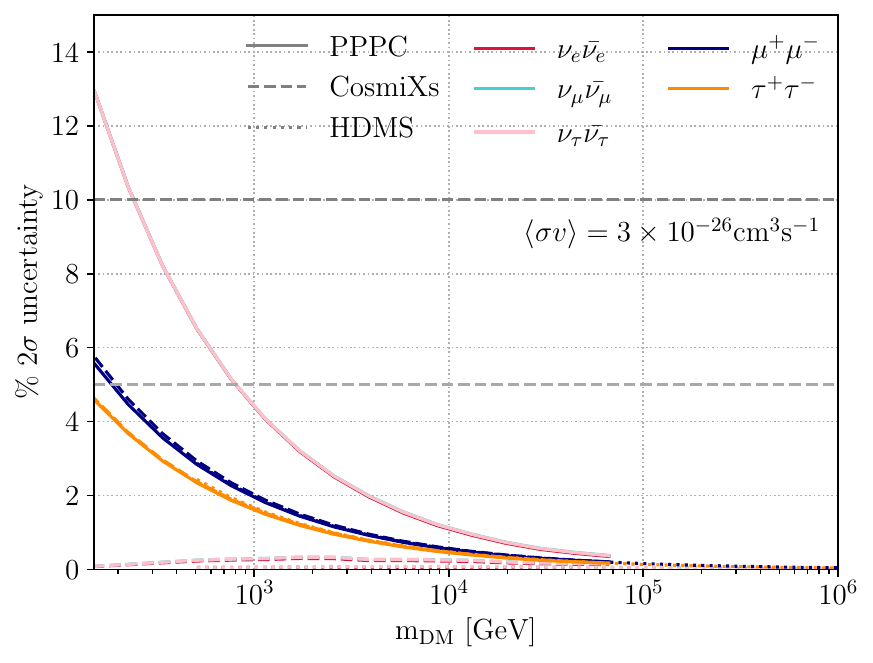}

    \caption{\footnotesize{$2\sigma$ flux uncertainty needed for a future neutrino telescope to be sensitive to a decay lifetime of $\tau=10^{29}$~s (left panel) and the thermal relic cross-section $\langle \sigma v \rangle_\mathrm{th}$ (right panel). The results correspond to the \texttt{PPPC} (solid lines), \texttt{CosmiXs} (dashed lines) and \texttt{HDMS} (dotted lines). The horizontal grey lines represent the expected ideal maximum flux precision for KMN3NeT.}} 
\label{fig:all_uncertainty_CosmiXs_HDMS} 
\end{figure*}

Also, extending Fig.~\ref{fig:all_uncertainty}, in Fig.~\ref{fig:all_uncertainty_CosmiXs_HDMS} we show the uncertainty needed over the best fit to distinguish in the neutrino signal the DM flux over the astrophysical background (left panel for decaying DM and right panel for the annihilating case). Similar results are obtained with the three codes (solid lines for \texttt{PPPC}, dashed for \texttt{CosmiXs} and dotted for \texttt{HDMS}), except in the case of the neutrino channels in the annihilating case. The difference is because the peak of the flux is smaller in these last two codes than in \texttt{PPPC}, with the consequence that a greater precision is needed for the other two codes.

\section{Constraints comparison in the $\tau^+\tau^-$ channel}
\label{ap:app_tau_constraints_comparison}

\begin{figure*}[t!] 
    \centering 
    \includegraphics[width=0.46\textwidth]{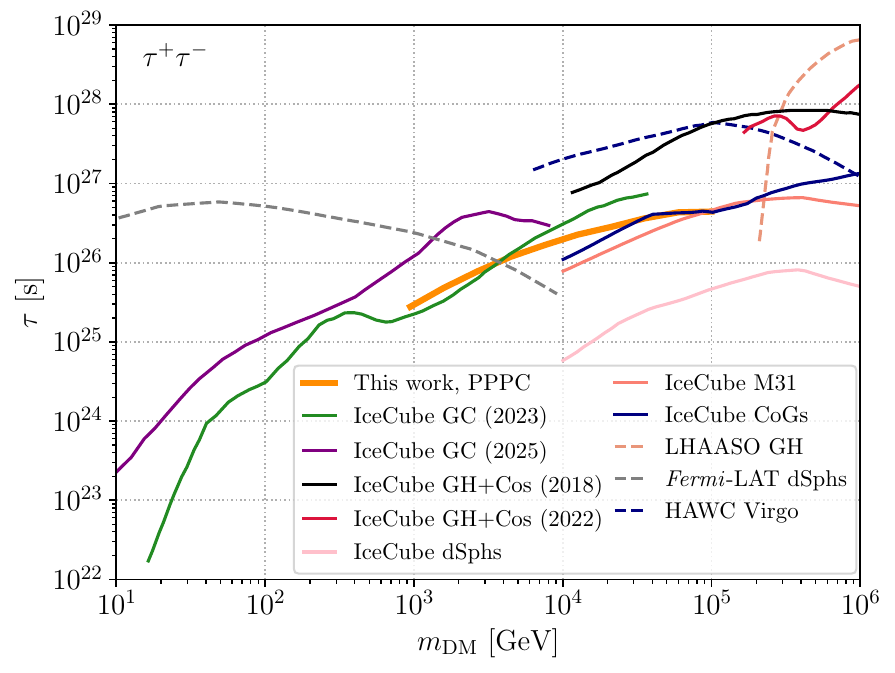}
    \includegraphics[width=0.46\textwidth]{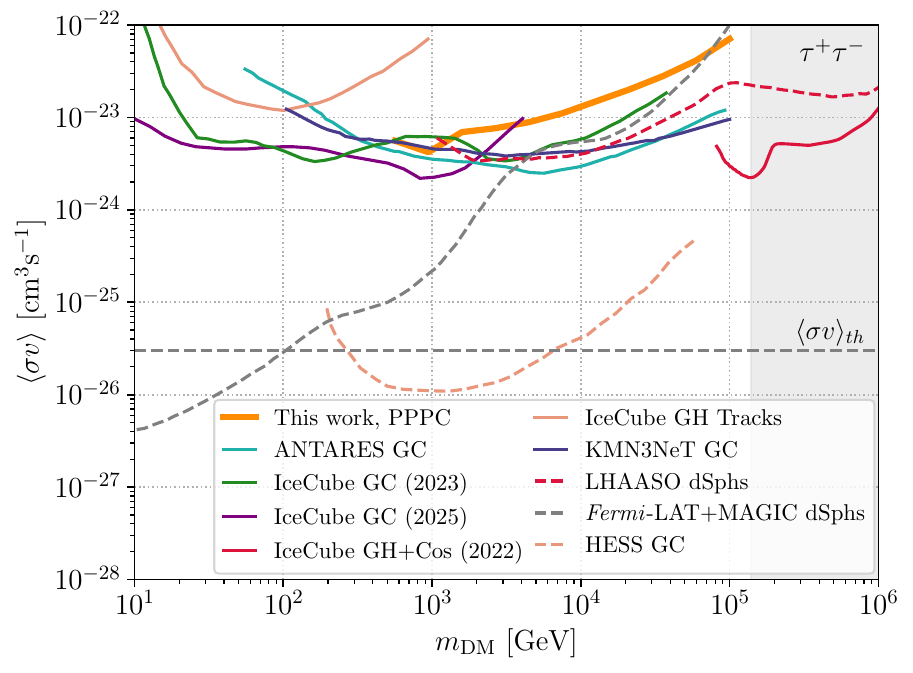}
    \caption{\footnotesize{$90\%$ C.L. lower limits on the decay lifetime $\tau$ (left panel) and upper limits on the annihilation cross-section $\langle \sigma v \rangle$ (right panel) for the $\tau^+\tau^-$ channel. Our results are shown with the thick yellow (\texttt{PPPC}, assuming the NFW profile) lines, whereas the solid lines are for other neutrino constraints and the dashed lines are for gamma-ray telescopes. References: IceCube GC (2023) \cite{2023PhRvD.108j2004A}, IceCube GC (2025) \cite{2025arXiv251100918T}, IceCube GH+Cos (2018) \cite{2018EPJC...78..831A}, IceCube GH+Cos (2022) \cite{2023JCAP...10..003A}, IceCube dSphs, M31 and CoGs \cite{2023arXiv230804833J}, LHAASO GH \cite{2022PhRvL.129z1103C}, \textit{Fermi}-LAT+MAGIC \cite{2016JCAP...02..039M}, HAWC Virgo \cite{2024PhRvD.109d3034A}, ANTARES GC \cite{ALBERT2020135439}, IceCube GH Tracks \cite{2017EPJC...77..627A}, KMN3Net GC \cite{2025JCAP...03..058A}, LHAASO dSphs \cite{2024PhRvL.133f1001C} and HESS GC \cite{PhysRevLett.129.111101}. The grey band in the bottom panel indicates when the unitarity limit breaks for thermal cold DM ($m_\mathrm{DM} \gtrsim 140$ TeV).}}
\label{fig:tau_constraints_comparison} 
\end{figure*}

In this Appendix, we show in Fig.~\ref{fig:tau_constraints_comparison} the comparison of our constraints (yellow thick line, \texttt{PPPC} assuming the NFW profile) with the rest of the literature, with limits computed from neutrino \cite{2018EPJC...78..831A, 2023JCAP...10..003A, 2023PhRvD.108j2004A, 2025arXiv251100918T, 2023arXiv230804833J, ALBERT2020135439, 2017EPJC...77..627A} (solid lines) and gamma-ray telescopes \cite{2022PhRvL.129z1103C, 2024PhRvD.109d3034A, 2024PhRvL.133f1001C, 2016JCAP...02..039M, PhysRevLett.129.111101} (dashed lines). For decay DM (left panel), our constraints are competitive with other neutrino constraints at lower masses, but at higher masses, IceCube's constraints are 1 order of magnitude better. For the annihilating case (right panel), our constraints are competitive compared to IceCube, ANTARES, or KMN3NeT. In all cases, neutrino constraints are in general several orders of magnitude less constraining than the gamma-ray limits. The grey band in the bottom panel indicates when the unitarity limit breaks for thermal cold DM ($m_\mathrm{DM} \gtrsim 140$ TeV).

\section{Constraints on the Yukawa couplings and on the individual decay channels}
\label{ap:app_ind_constraints_neutrinoful}

\begin{figure*}[t!] 
    \centering 
    \includegraphics[width=0.46\textwidth]{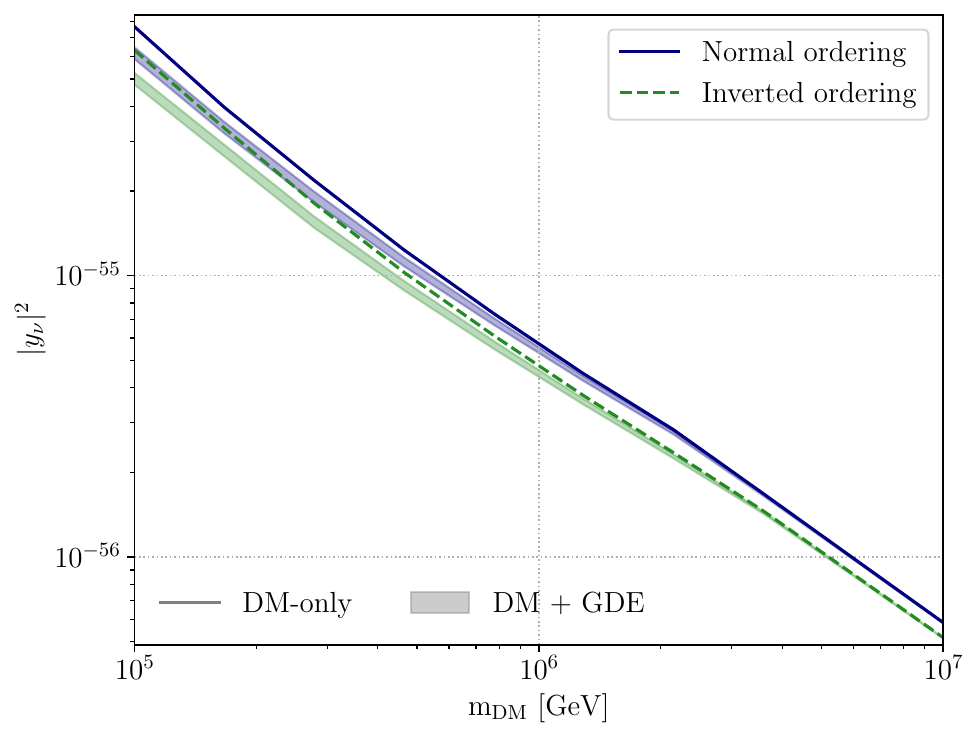}
    \includegraphics[width=0.46\textwidth]{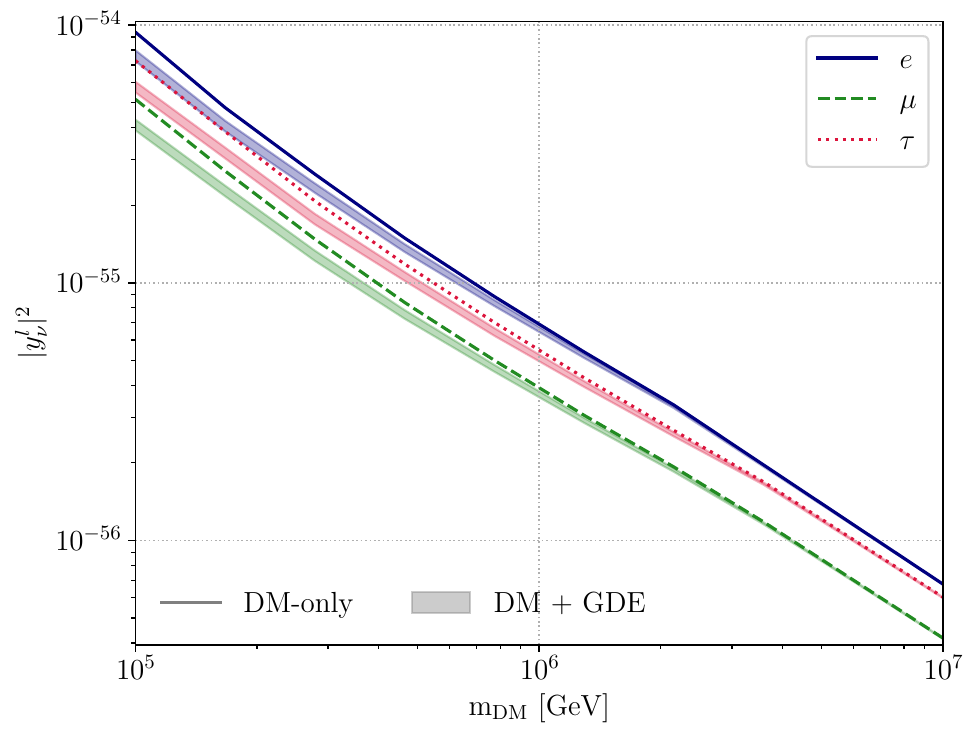}
    \caption{\footnotesize{Same as Fig.~\ref{fig:neutrinoful}, but we show the upper limits on the Yukawa couplings of each of the cases considered.}}
\label{fig:yukawas_neutrinoful} 
\end{figure*}

\begin{figure*}[t!] 
    \centering 
    \includegraphics[width=0.99\textwidth]{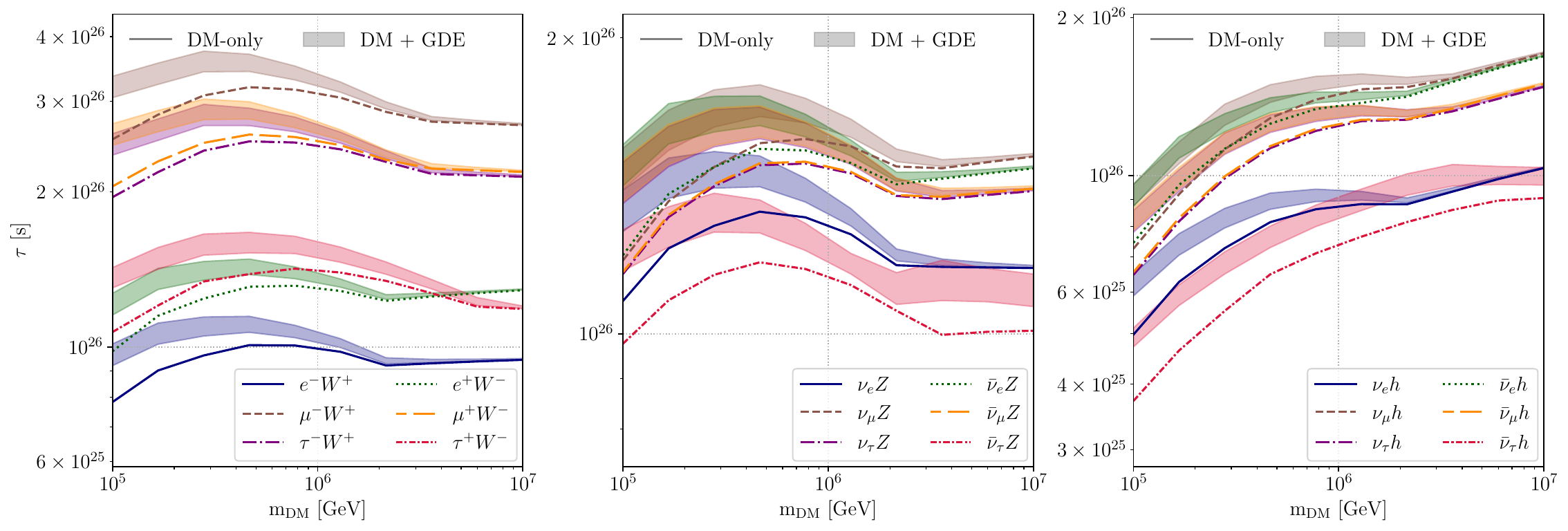}
    \includegraphics[width=0.99\textwidth]{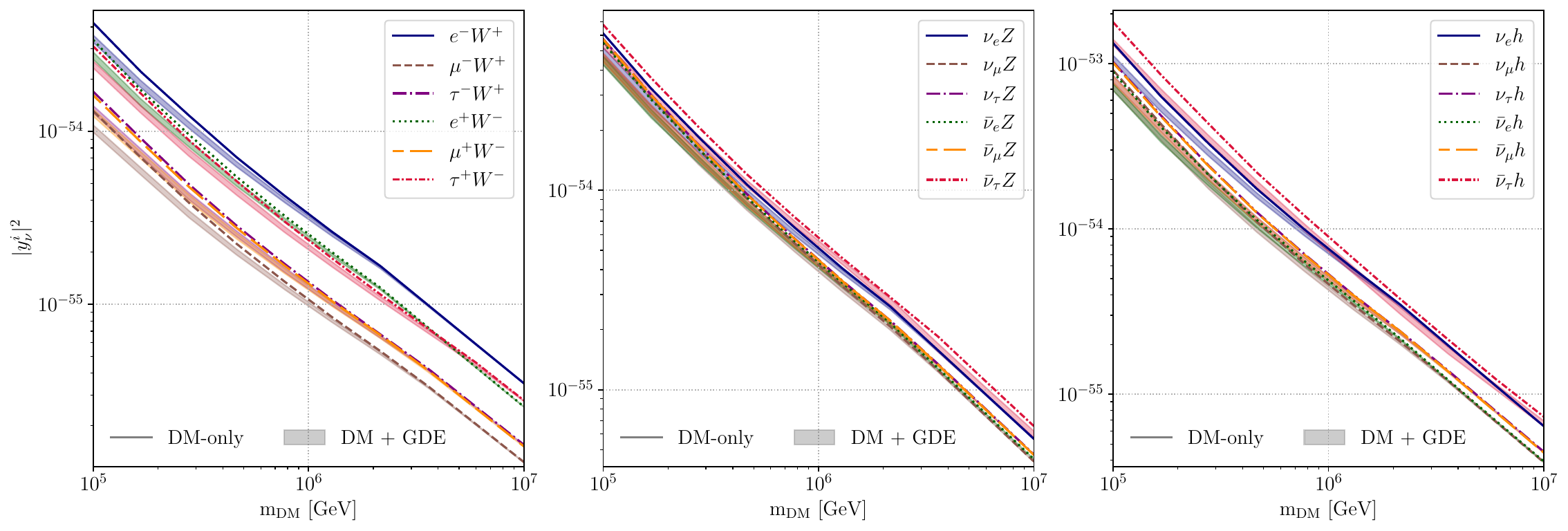}
    \caption{\footnotesize{Lower limits on the decay lifetime $\tau$ (first row) and upper limits on the specific Yukawa coupling $|y^i_\nu|$ (second row), assuming only one individual decay channel. These constraints are computed assuming \texttt{HDMS} and the NFW profile.}}
\label{fig:ind_channel_neutrinoful} 
\end{figure*}

In Fig.~\ref{fig:yukawas_neutrinoful}, similarly to Fig.~\ref{fig:neutrinoful}, we show the corresponding upper limits on the Yukawa couplings for the complete model (left panel) and for each of the leptonic channels (right panel). Finally, to investigate a case-by-case scenario, Fig.~\ref{fig:ind_channel_neutrinoful} illustrates the individual constraints assuming only one decay channel $i$, depicted by the decay lifetime $\tau$ lower limits (first row) and the specific Yukawa coupling $|y^i_\nu|$ upper limits (second row).

\newpage
\bibliographystyle{else-num}
\bibliography{biblio}
\end{document}